\documentclass[11pt]{article}
\usepackage{mathrsfs}
\usepackage{amsfonts}
\usepackage{amsmath}
\usepackage{amssymb}

\numberwithin{equation}{section}

\author{ Yu Hou   \and  Engui Fan\footnote{Corresponding
author and  e-mail address:
      faneg@fudan.edu.cn} \and  Peng Zhao}
\date{   \small{ School of Mathematical Sciences, Institute of Mathematics \\
 and Key Laboratory of Mathematics for Nonlinear Science, \\ Fudan
University, Shanghai 200433, P.R. China}}
\title{\bf \Large{The algebro-geometric solutions for Hunter-Saxton hierarchy} }
\begin{document}
\maketitle
\begin{abstract}
This paper is dedicated to provide theta function representation of
algebro-geometric solutions and related crucial quantities for the
Hunter-Saxton (HS) hierarchy through studying a algebro-geometric
initial value problem. Our main tools include the polynomial
recursive formalism to derive the HS hierarchy, the hyperelliptic
curve with finite number of genus, the Baker-Akhiezer functions, the
meromorphic function, the Dubrovin-type equations for auxiliary
divisors, and the associated trace formulas. With the help of these
tools, the explicit representations of the Baker-Ahhiezer functions,
the meromorphic function, and the algebro-geometric solutions  are
obtained for the entire HS hierarchy.
\end{abstract}

\section{Introduction}
 The Hunter-Saxton (HS) equation
     \begin{equation}\label{1.1}
        u_{xxt}=-2uu_{xxx}-4u_{x}u_{xx},
     \end{equation}
 where $u(x,t)$ is the function of spatial variable $x$ and time
 variable $t$.
 It arises in two different physical contexts in two nonequivalent
 variational forms \cite{1}, \cite{2}. The first is shown to describe the propagation
 of weakly nonlinear orientation waves in a massive nematic liquid
 crystal director field \cite{1}-\cite{3}. The second is shown to describe the high
 frequency limit of the Camassa-Holm (CH) equation \cite{5},
 \cite{6}, \cite{32}
   \begin{equation}\label{1.2}
   u_t-u_{xxt}+3uu_x=2u_xu_{xx}+uu_{xxx}
   \end{equation}
 which was originally introduced in \cite{5}, \cite{6} as model equation for shallow
 water waves, and obtained independently in \cite{31} with a bi-hamiltonian
 structure.

 The HS equation is a completely integrable system with a bi-hamiltonian
 structure and hence it possesses a Lax pair, an infinite family of commuting
 Hamiltonian flows, as well as an associated sequence of conservation laws,
 (Hunter and Zheng \cite{2}, Reyes \cite{20}). The inverse scattering
 solutions have been obtained by Beals, Sattinger and Szmigielski \cite{19}.
 Recently, Lenells \cite{23}, \cite{24} and also Khesin and Misio{\l}ek \cite{22} pointed
 out that it describes the geodesic flow on the homogeneous space related to
 the Virasoro group. Bressan and Constantin \cite{25}, also Holden \cite{26}
 constructed a continuous semigroup
 of weak, dissipative solutions. Yin \cite{27} proved the local existence of strong
 solutions of the periodic HS equation and showed that all strong solutions-except
 space independent solutions-blow up in finite time. Gui, Liu and
 Zhu \cite{28} studied the wave-breaking phenomena and global existence.
 Furthermore, Morozov \cite{29}, Sakovich \cite{30} and Reyes \cite{20}, \cite{21}
 investigated (\ref{1.1}) from
 a geometric perspective. However, within the knowledge of the authors,
 the algebro-geometric solutions of the entire HS hierarchy are not studied yet.

 The main task of this paper focuses on the algebro-geometric solutions
 of the whole HS hierarchy in which (\ref{1.1}) is just the first member.
 Algebro-geometric solution, an important feature of integrable system,
 is a kind of explicit solutions closely related to the inverse spectral
 theory \cite{4}, \cite{7}, \cite{9}-\cite{11}. As a degenerated case of the
 algebro-geometric solution, the multi-soliton solution and periodic solution
 in elliptic function type may be obtained \cite{7}, \cite{8}, \cite{33}.
 A systematic approach, proposed by Gesztesy and Holden to construct
 algebro-geometric solutions for integrable equations,   has been extended to
 the whole (1+1) dimensional integrable hierarchy, such as the AKNS hierarchy,
 the CH hierarchy etc. \cite{12}-\cite{15}. Recently, we investigated algebro-geometric solutions for the
 Gerdjikov-Ivanov hierarchy, the Degasperis-Procesi hierarchy and
 the modified Camassa-Holm hierarchy \cite{16}-\cite{18}.

 The outline of the present paper is as follows.

 In section 2, based
 on the polynomial recursion formalism, we derive the HS hierarchy,
 the associated sequences, and Lax pairs. A hyperelliptic curve
 $\mathcal{K}_{n}$ with arithmetic genus $n$ is introduced with
 the help of the characteristic polynomial of Lax matrix $V_n$ for the
 stationary HS hierarchy.

 In Section 3, we study a meromorphic
 function $\phi$ such that $\phi$ satisfies a nonlinear
 second-order differential equation. Then we study the properties of
 the Baker-Akhiezer function $\psi$, and furthermore the stationary
 HS equations are decomposed into a system of Dubrovin-type
 equations. The stationary trace formulas
 are obtained for the HS hierarchy.

 In Section 4, we present the first set of our results, the explicit
 theta function representations of Baker-Akhiezer function, the
 meromorphic function  and the potentials $u$
 for the entire stationary HS hierarchy. Furthermore, we study the
 initial value problem on an algebro-geometric curve for the stationary
 HS hierarchy.

 In Sections 5 and 6, we extend the analysis in Sections 3 and 4
 to the time-dependent case. Each equation in the HS
 hierarchy is permitted to evolve in terms of an independent time
 parameter $t_r$. As an initial data we use a stationary solution of
 the $n$th equation and then construct a time-dependent solution of
 the $r$th equation in the HS hierarchy.
 The Baker-Akhiezer function, the meromorphic function, the analogs
 of the Dubrovin-type equations, the trace formulas, and the theta function
 representation in Section 4 are all extended to the time-dependent
 case.

\section{The HS hierarchy}
 In this section, we derive the HS hierarchy and the corresponding
 sequence of zero-curvature pairs by using a polynomial recursion formalism.
 Moreover, we introduce the hyperelliptic curve connecting to
 the stationary HS hierarchy.

 Throughout this section, let us we make the following hypothesis.

\newtheorem{hyp1}{Hypothesis}[section]
 \begin{hyp1}
      In the stationary case we assume that
  \begin{equation}\label{2.1}
    \begin{split}
    & u\in C^\infty(\mathbb{R}),\ \
   \partial_x^ku\in L^\infty (\mathbb{R}),\qquad k\in \mathbb{N}_0.
    \end{split}
  \end{equation}
    In the time-dependent case we suppose
  \begin{equation}
    \begin{split}\label{2.2}
    & u(\cdot,t)\in C^\infty (\mathbb{R}),\ \
     \partial_x^ku(\cdot,t)\in L^\infty
    (\mathbb{R}),\quad k\in \mathbb{N}_0,~~ t\in \mathbb{R},\\
    & u(x,\cdot), u_{xx}(x,\cdot)\in C^1(\mathbb{R}),
       \quad x\in \mathbb{R}.\\
  \end{split}
  \end{equation}
\end{hyp1}

We start by the polynomial recursion formalism. Define
 $\{f_{ l}\}_{l\in\mathbb{N}_{0}}$,
 $\{g_{ l}\}_{l\in\mathbb{N}_{0}}$ and
 $\{h_{ l}\}_{l\in\mathbb{N}_{0}}$
recursively by
    \begin{equation}\label{2.3}
      \begin{split}
        & f_0=1, \\
       & f_{l+1,x}=\mathcal{G}(-4u_{xx}f_{l,x}-2u_{xxx}f_l),
        \quad l \in \mathbb{N}_0,\\
      & g_l=\frac{1}{2}f_{l+1,x},\quad l \in \mathbb{N}_0,\\
      & h_l=-g_{l+1,x}-u_{xx}f_{l+1}, \quad l \in \mathbb{N}_0,\\
      \end{split}
    \end{equation}
where $\mathcal{G}$ is given by
    \begin{equation}\label{2.4}
     \begin{split}
     & \mathcal{G}: L^\infty(\mathbb{R}) \rightarrow
        L^\infty(\mathbb{R}),\\
     & (\mathcal{G}v)(x)=\int_{-\infty}^x \int_{-\infty}^\tau
     v(y) ~dy d\tau , \quad x\in\mathbb{R},~
        v\in L^\infty(\mathbb{R}).
      \end{split}
    \end{equation}
It is easy to see that $\mathcal{G}$ is the resolvent of the
one-dimensional Laplacian operator, that is
    \begin{equation}\label{2.5}
        \mathcal{G}=\Big( \frac{d^2}{dx^2} \Big)^{-1}.
    \end{equation}
Explicitly, one computes
   \begin{equation}\label{2.6}
     \begin{split}
      & f_0=1,\\
      & f_1=-2u+c_1,\\
      & f_2=\mathcal{G}(4uu_{xx}+2u_x^2)-c_1 2u +c_2,\\
      & g_0=-u_x,\\
      & g_1=\frac{1}{2} \mathcal{G}(8u_xu_{xx}+4uu_{xxx})-c_1u_x,\\
      & h_0=-\frac{1}{2}f_{2,xx}-u_{xx}f_1,
     \end{split}
   \end{equation}
where $\{c_l\}_{l\in\mathbb{N}_0}\subset\mathbb{C}$ are integration
constants and we have used the assumption
   \begin{equation}\label{2.7}
     f_l(u)|_{u=0}=c_l, ~~g_l(u)|_{u=0}=c_l,
   ~~h_l(u)|_{u=0}=c_l, ~~l \in \mathbb{N}.
   \end{equation}

Next we introduce the corresponding homogeneous coefficients
$\hat{f}_{l}, \hat{g}_{l},$ and  $\hat{h}_{l},$ defined through
taking $c_k=0$ for $k=1,\cdots,l,$
    \begin{equation}\label{2.8}
       \begin{split}
        & \hat{f}_0=f_0=1, \quad
        \quad \hat{f}_{l}=f_{l}|_{c_k=0,~k=1,\ldots,l},
         \quad l \in \mathbb{N},\\
        & \hat{g}_0=g_0=-u_x, \quad
        \hat{g}_{l}=g_{l}|_{c_k=0,~k=1,\ldots,l},
        \quad  l \in \mathbb{N}.\\
        &   \hat{h}_0=h_0, \quad
           \quad \hat{h}_{l}=h_{l}|_{c_k=0,~k=1,\ldots,l},
           \quad l \in \mathbb{N}.
       \end{split}
    \end{equation}
Hence one can easily conclude that
    \begin{equation}\label{2.9}
      f_{l}=\sum_{k=0}^{l}c_{l-k}\hat{f}_{k}, \quad
      g_{l}=\sum_{k=0}^{l}c_{l-k}\hat{g}_{k}, \quad
      h_{l}=\sum_{k=0}^{l}c_{l-k}\hat{h}_{k}, \quad
       l\in \mathbb{N}_0,
    \end{equation}
with
    \begin{equation}\label{2.10}
        c_0=1.
    \end{equation}
Now we consider the following  $2\times 2$ matrix isospectral
problem
     \begin{equation}\label{2.11}
       \psi_x=U(u,z)\psi=
       \left(
         \begin{array}{cc}
           0 & 1 \\
           -z^{-1}u_{xx} & 0 \\
         \end{array}
       \right)
       \psi
     \end{equation}
and an auxiliary problem
    \begin{equation} \label{2.12}
       \psi_{t_n}=V_n(z)\psi,
    \end{equation}
where $V_n(z)$ is defined by
    \begin{equation}\label{2.13}
      V_n(z)=
      \left(
        \begin{array}{cc}
          -G_n(z) & F_{n+1}(z) \\
          z^{-1}H_n(z) & G_n(z) \\
        \end{array}
      \right)
    \qquad z \in \mathbb{C} \setminus \{0\},
    \quad n\in\mathbb{N}_0,
    \end{equation}
assuming $F_{n+1}$, $G_n$ and $H_n$ to be polynomials of degree $n$
with $C^\infty$ coefficients with respect to $x$. The compatibility
condition between  (\ref{2.11}) and (\ref{2.12}) yields the
stationary zero-curvature equation
\begin{equation}\label{2.14}
   -V_{n,x}+[U,V_n]=0,
\end{equation}
that is
   \begin{eqnarray}\label{2.15}
      F_{n+1,x}&=& 2G_n, \\
      H_{n,x}&=& 2u_{xx}G_n, \\
      zG_{n,x}&=&-H_n -u_{xx}F_{n+1}.
    \end{eqnarray}
From (\ref{2.15})-(2.17), a direct calculation shows that
   \begin{equation}\label{2.18}
     \frac{d}{dx} \mathrm{det} (V_n(z,x))=
     -\frac{1}{z^2} \frac{d}{dx} \Big(
     z^2G_n(z,x)^2+zF_{n+1}(z,x)H_n(z,x)
     \Big)=0
   \end{equation}
and hence $z^2G_n^2+zF_{n+1}H_n$ is $x$-independent implying
   \begin{equation}\label{2.19}
    z^2G_n^2+zF_{n+1}H_n=R_{2n+2},
   \end{equation}
where the integration constant $R_{2n+2}$ is a polynomial of degree
$2n+2$ with respect to $z$. If $\{E_m\}_{m=0,\cdots,2n+1}$ denote
its zeros, then
   \begin{equation}\label{2.20}
   R_{2n+2}(z)=(u_x^2+h_0)\prod_{m=0}^{2n+1}(z-E_m),\quad
   E_0=0,~\{E_m\}_{m=1,\cdots,2n+1}\in\mathbb{C}.
   \end{equation}
Here we must emphasize that the coefficient $(u_x^2+h_0)$ is a
constant. In fact, (\ref{2.18}) equals
    \begin{equation}\label{2.21}
     2z^2G_nG_{n,x}+zF_{n+1}H_{n,x}+zH_nF_{n+1,x}=0.
    \end{equation}
Comparing the coefficients of powers $z^{2n+2}$ yields
   \begin{equation}\label{2.22}
    2g_0g_{0,x}+f_0h_{0,x}+h_0f_{0,x}=0,
   \end{equation}
which together with (\ref{2.6}) we obtain
    \begin{equation}\label{2.23}
        2u_xu_{xx}+h_{0,x}=0.
    \end{equation}
Hence
    \begin{equation}\label{2.24}
        u_x^2+h_0=\partial^{-1} (2u_xu_{xx}+h_{0,x})=
        \mathrm{Constant}.
    \end{equation}
For simplicity, we denote it by $a^2$. Then $R_{2n+2}(z)$ can be
rewritten as
    \begin{equation}\label{2.25}
   R_{2n+2}(z)=a^2 \prod_{m=0}^{2n+1}(z-E_m),\quad
   E_0=0,~\{E_m\}_{m=1,\cdots,2n+1}\in\mathbb{C}.
   \end{equation}
In order to derive the corresponding hyperelliptic curve, we compute
the characteristic polynomial $\mathrm{det}(yI-z V_n)$ of Lax matrix
$z V_n$,
    \begin{eqnarray}\label{2.26}
      \mathrm{det}(yI-z V_n)&=&
        y^2-z^2G_n(z)^2-F_{n+1}(z)H_n(z)
         \nonumber \\
      &=& y^2-R_{2n+2}(z)=0.
    \end{eqnarray}
Equation (\ref{2.26}) naturally leads to the hyperelliptic curve
$\mathcal{K}_n$, where
  \begin{equation}\label{2.27}
     \mathcal {K}_n:\mathcal {F}_n(z,y)=y^2-R_{2n+2}(z)=0.
   \end{equation}

 The stationary zero-curvature equation
(\ref{2.14}) implies polynomial recursion relations (\ref{2.3}).
Introducing the following polynomial $F_{n+1}(z), G_n(z)$ and
$H_n(z)$ with respect to the spectral parameter $z$,
   \begin{equation}\label{2.28}
     F_{n+1}(z)=\sum_{l=0}^{n+1} f_{l} z^{n+1-l},
   \end{equation}

   \begin{equation}\label{2.29}
     G_n(z)=\sum_{l=0}^n g_{l} z^{n-l},
   \end{equation}

   \begin{equation}\label{2.30}
     H_n(z)=\sum_{l=0}^n h_{l} z^{n-l}.
   \end{equation}
Inserting (\ref{2.28})-(\ref{2.30}) into (2.15)-(2.17) then yields
the recursions relations (\ref{2.3}) for $f_{l},$ $l=0,\ldots, n+1,$
and $g_{l}$, $l=0,\ldots, n.$  For fixed $n \in \mathbb{N}_0$, by
using (2.17), we obtain the recursion for $h_{l}, $ $l=0, \ldots,
n-1$ in (\ref{2.3}) and
    \begin{equation}\label{2.31}
        h_{n}=-u_{xx}f_{n+1}.
     \end{equation}
Moreover, from (2.16), one infers that
   \begin{equation}\label{2.32}
      h_{n,x}-2u_{xx}g_{n}=0,
      \qquad n\in \mathbb{N}_0.
   \end{equation}
Hence, insertion of  the equation (\ref{2.31}) and
     \begin{equation}\label{2.33}
        f_{n+1,x}-2g_{n}=0
     \end{equation}
into (\ref{2.32}), we derive the stationary HS hierarchy,
      \begin{equation}\label{2.34}
        \textrm{s-HS}_n(u)=
        u_{xxx}f_{n+1}(u)+2u_{xx}f_{n+1,x}(u)=0,
        \quad n\in \mathbb{N}_0.
      \end{equation}
Explicitly, the first few equations are as follows
    \begin{equation}\label{2.35}
      \begin{split}
       & \textrm{s-HS}_0(u)=-2uu_{xxx}-4u_xu_{xx}+c_1u_{xxx}=0,\\
       & \textrm{s-HS}_1(u)=u_{xxx} \mathcal{G}(4uu_{xx}+2u_x^2)
       +2u_{xx} \mathcal{G}(8u_xu_{xx}+4uu_{xxx})\\
       & ~~~~~~~~~~~~~~
          +c_1(-2uu_{xxx}-4u_xu_{xx})+c_2u_{xxx}=0,\\
       & \mathrm{etc}.
      \end{split}
    \end{equation}
By definition, the set of solutions of (\ref{2.34}) represents the
class of algebro-geometric HS solutions, with $n$ ranging in
$\mathbb{N}_0$ and $c_l$ in $\mathbb{C},~l\in\mathbb{N}$. We call
the stationary algebro-geometric HS solutions $u$ as HS potentials
at times.

\newtheorem{rem2.2}[hyp1]{Remark}
  \begin{rem2.2}
    Here we emphasize that if $u$ satisfies one of the stationary HS equations in
    $(\ref{2.34})$, then it must satisfy infinitely many such equations
    of order higher than $n$ for certain choices of integration
    constants $c_l$, this is a common characteristic of the general
    integrable soliton equations such as the KdV, AKNS and CH
    hierarchies \cite{15}.
  \end{rem2.2}

Next, we introduce the corresponding homogeneous polynomials
$\widehat{F}_{l+1}, \widehat{G}_{l}, \widehat{H}_{l}$ defined by
    \begin{eqnarray}\label{2.36}
    &&
      \widehat{F}_{l+1}(z)=F_{l+1}(z)|_{c_k=0,~k=1,\dots,l}
      =\sum_{k=0}^{l+1} \hat{f}_{k} z^{l+1-k},
      ~~ l=0,\ldots,n,\\
    &&
      \widehat{G}_{l}(z)=G_{l}(z)|_{c_k=0,~k=1,\dots,l}
      =\sum_{k=0}^l \hat{g}_{k} z^{l-k},
      ~~ l=0,\ldots,n,\\
    &&
      \widehat{H}_{l}(z)=H_{l}(z)|_{c_k=0,~k=1,\dots,l}
      =\sum_{k=0}^l \hat{h}_{k} z^{l-k},
      ~~ l=0,\ldots,n-1,\\
    &&
      \widehat{H}_{n}(z)= -u_{xx}\hat{f}_{n+1}+
      \sum_{k=0}^{n-1} \hat{h}_{k} z^{n-k}.
          \end{eqnarray}
Then the corresponding homogeneous formalism of (\ref{2.34}) are
given by
      \begin{equation}\label{2.40}
        \textrm{s-}\widehat{\mathrm{HS}}_n(u)=
        \textrm{s-HS}_n(u)|_{c_l=0,~l=1,\dots,n}=0,
        \qquad n\in\mathbb{N}_0.
      \end{equation}

 We will end this section by introducing the time-dependent HS hierarchy. This means that $u$
are now considered as functions of both space and time. We introduce
a deformation parameter $t_n \in \mathbb{R}$ in $u$, replacing
$u(x)$ by $u(x,t_n)$, for each equation in the hierarchy. In
addition, we note that the definitions (\ref{2.11}), (\ref{2.13})
and (2.28)-(2.30) of $U,$ $V_n$ and $F_{n+1}, G_n$ and $H_n$ are
still apply. Then the compatibility condition yields the
zero-curvature equation
   \begin{equation}\label{2.41}
   U_{t_n}-V_{n,x}+[U,V_n]=0, \qquad n\in \mathbb{N}_0,
   \end{equation}
namely
    \begin{eqnarray}
      && -u_{xxt_n}-H_{n,x}+2u_{xx}G_n=0, \\
      && F_{n+1,x}=2G_n, \\
      && zG_{n,x}=-H_n -u_{xx}F_{n+1}.
    \end{eqnarray}
For fixed $n \in \mathbb{N}$, insertion of the polynomial
expressions for $F_{n+1}$, $G_n$ and $H_n$ into (2.42)-(2.44),
respectively, then we derive the relations (\ref{2.3}) for
$f_{l}|_{l=0,\ldots,n+1}$, $g_{l}|_{l=0,\ldots,n}$.
$h_{l}|_{l=0,\ldots,n-1}$ and
      \begin{equation}\label{2.45}
        h_{n}=-u_{xx}f_{n+1}.
      \end{equation}
Moreover, from (2.42), we infer that
     \begin{equation}\label{2.46}
        -u_{xxt_n}-h_{n,x}+2u_{xx}g_n=0,
        \qquad n\in \mathbb{N}_0.
     \end{equation}
Hence, together (\ref{2.45}) and
      \begin{equation}\label{2.47}
        f_{n+1,x}=2g_n,
      \end{equation}
(\ref{2.46}) admits the time-dependent HS hierarchy,
     \begin{eqnarray}\label{2.48}
        \mathrm{HS}_n(u)=
        -u_{xxt_n}+u_{xxx}f_{n+1}(u)+2u_{xx}f_{n+1,x}(u)=0,
         \nonumber \\
        \quad (x,t_n)\in \mathbb{R}^2,
        ~ n\in\mathbb{N}_0.
     \end{eqnarray}
Explicitly, the first few equations are as follows
     \begin{equation}\label{2.49}
      \begin{split}
       &
       \mathrm{HS}_0(u)=-u_{xxt_0}-2uu_{xxx}-4u_{xx}u_{x}+c_1u_{xxx}
                       =0,\\
       & \mathrm{HS}_1(u)=-u_{xxt_1}+ u_{xxx} \mathcal{G}(4uu_{xx}+2u_x^2)
                        +2u_{xx} \mathcal{G}(8u_xu_{xx}+4uu_{xxx})
                        \\
       & ~~~~~~~~~~~~
           + c_1(-2uu_{xxx}-4u_xu_{xx})+c_2u_{xxx}=0,\\
       & \mathrm{etc}.
      \end{split}
     \end{equation}
The first equation $\mathrm{HS}_0(u)=0$ (with $c_1=0$) in the
hierarchy represents the Hunter-Saxton equation discussed in section
1. Similarly, one can introduce the corresponding homogeneous HS
hierarchy by
    \begin{equation}\label{2.50}
        \widehat{\mathrm{HS}}_n(u)=\mathrm{HS}_n(u)|_{c_l=0,~l=1,\ldots,n}=0,
        \qquad n\in\mathbb{N}_0.
    \end{equation}

In fact, since the Lenard recursion formalism is almost universally
adopted in the contemporary literature on the integrable soliton
equations, it might be worthwhile to adopt Gesztesy method, an
alternative approach using the polynomial recursion relations.

\section{The stationary HS formalism}
 In this section we focus our attention on the stationary case. By
 using the polynomial recursion formalism described in section 2, we
 define a fundamental meromorphic function $\phi(P,x)$ on a hyperelliptic
 curve $\mathcal{K}_n$. Moreover, we study the properties of the
 Baker-Akhiezer function $\psi(P,x,x_0)$, Dubrovin-type equations
 and trace formulas.

 We emphasize that the analysis about the stationary case described in section
 2 also holds here for the present context.

 The hyperelliptic curve $\mathcal{K}_n$
     \begin{equation}\label{3.1}
       \begin{split}
        & \mathcal{K}_n:  \mathcal{F}_n(z,y)=y^2-R_{2n+2}(z)=0, \\
        & R_{2n+2}(z)=a^2 \prod_{m=0}^{2n+1} (z-E_m), \quad
          E_0=0,~ \{E_m\}_{m=1,\ldots,2n+1} \in \mathbb{C},
        \end{split}
     \end{equation}
which is compactified by joining two points at infinity,
$P_{\infty_\pm}$, $P_{\infty_+} \neq P_{\infty_-}$, but for
notational simplicity the compactification is also denoted by
$\mathcal{K}_n$. Points $P$ on
      $$\mathcal{K}_{n} \setminus \{P_{\infty_+},P_{\infty_-}\}$$
are represented as pairs $P=(z,y(P))$, where $y(\cdot)$ is the
meromorphic function on $\mathcal{K}_{n}$ satisfying
       $$\mathcal{F}_n(z,y(P))=0.$$
The complex structure on $\mathcal{K}_{n}$ is defined in the usual
way by introducing local coordinates
$$\zeta_{Q_0}:P\rightarrow(z-z_0)$$
near points $Q_0=(z_0,y(Q_0))\in \mathcal{K}_{n} \setminus P_0, ~
P_0=(0,0),$ which are neither branch nor singular points of
$\mathcal{K}_{n}$; near $P_0$, the local coordinates are
   $$\zeta_{P_0}:P \rightarrow z^{1/2};$$
near the points $P_{\infty_\pm} \in \mathcal{K}_{n}$, the local
coordinates are
   $$\zeta_{P_{\infty_\pm}}:P \rightarrow z^{-1},$$
and similarly at branch and singular points of $\mathcal{K}_{n}.$
Hence, $\mathcal{K}_{n}$ becomes a two-sheeted hyperelliptic Riemann
surface of genus $n \in \mathbb{N}_0$ (possibly with a singular
affine part) in a standard manner.

We also notice that fixing the zeros $E_0=0,$ $E_1,\ldots,E_{2n+1}$
of $R_{2n+2}$ discussed in (\ref{3.1}) leads to the curve
$\mathcal{K}_{n}$ is fixed. Then the integration constants
$c_1,\ldots,c_n$ in $f_{n}$ are uniquely determined, which is the
symmetric functions of $E_1,\ldots,E_{2n+1}$.

The holomorphic map
   $\ast,$ changing sheets, is defined by
       \begin{eqnarray}\label{3.2}
       && \ast: \begin{cases}
                        \mathcal{K}_{n}\rightarrow\mathcal{K}_{n},
                       \\
                       P=(z,y_j(z))\rightarrow
                       P^\ast=(z,y_{j+1(\mathrm{mod}~
                       2)}(z)), \quad j=0,1,
                      \end {cases}
                     \nonumber \\
      && P^{\ast \ast}:=(P^\ast)^\ast, \quad \mathrm{etc}.,
       \end{eqnarray}
where $y_j(z),\, j=0,1$ denote the two branches of $y(P)$ satisfying
$\mathcal{F}_{n}(z,y)=0$. \\
Finally, positive divisors on
$\mathcal{K}_{n}$ of degree $n$ are denoted by
        \begin{equation}\label{3.3}
          \mathcal{D}_{P_1,\ldots,P_{n}}:
             \begin{cases}
              \mathcal{K}_{n}\rightarrow \mathbb{N}_0,\\
              P\rightarrow \mathcal{D}_{P_1,\ldots,P_{n}}=
                \begin{cases}
                  \textrm{ $k$ if $P$ occurs $k$
                      times in $\{P_1,\ldots,P_{n}\},$}\\
                   \textrm{ $0$ if $P \notin
                     $$ \{P_1,\ldots,P_{n}\}.$}
                \end{cases}
             \end{cases}
        \end{equation}
Next, we define the stationary Baker-Akhiezer function
$\psi(P,x,x_0)$ on $\mathcal{K}_{n}\setminus \{P_{\infty_+},
P_{\infty_-},P_0\}$ by
       \begin{equation}\label{3.4}
         \begin{split}
          & \psi(P,x,x_0)=\left(
                            \begin{array}{c}
                              \psi_1(P,x,x_0) \\
                              \psi_2(P,x,x_0) \\
                            \end{array}
                          \right), \\
          & \psi_x(P,x,x_0)=U(u(x),z(P))\psi(P,x,x_0),\\
           & zV_n(u(x),z(P))\psi(P,x,x_0)=y(P)\psi(P,x,x_0),\\
          & \psi_1(P,x_0,x_0)=1; \\
           &   P=(z,y)\in \mathcal{K}_{n}
           \setminus \{P_{\infty_+},P_{\infty_-},P_0\},~(x,x_0)\in \mathbb{R}^2.
         \end{split}
       \end{equation}
Closely related to $\psi(P,x,x_0)$ is the following meromorphic
function $\phi(P,x)$ on $\mathcal{K}_{n}$ defined by
       \begin{equation}\label{3.5}
         \phi(P,x)= z
         \frac{ \psi_{1,x}(P,x,x_0)}{\psi_1(P,x,x_0)},
         \quad P\in \mathcal{K}_{n},~ x\in \mathbb{R}
       \end{equation}
with
   \begin{equation}\label{3.6}
    \psi_1(P,x,x_0)=\mathrm{exp}\left(z^{-1}\int_{x_0}^x
         \phi(P,x^\prime)~ dx^\prime
         \right),
         \quad P\in \mathcal{K}_{n}\setminus \{P_{\infty_+}, P_{\infty_-},
         P_0\}.
    \end{equation}
Then, based on (\ref{3.4}) and (\ref{3.5}), a direct calculation
shows that
    \begin{eqnarray}\label{3.7}
        \phi(P,x)&=&\frac{y+zG_n(z,x)}{F_{n+1}(z,x)}
           \nonumber \\
        &=&
         \frac{z H_n(z,x)}{y-zG_n(z,x)},
    \end{eqnarray}
and
      \begin{equation}\label{3.8}
        \psi_2(P,x,x_0)= \psi_1(P,x,x_0)\phi(P,x)/z.
      \end{equation}

We note that $F_{n+1}$ and $H_n$ are polynomials with respect to $z$
of degree $n+1$ and $n$, respectively. Hence we may write
        \begin{equation}\label{3.9}
         F_{n+1}(z)=\prod_{j=0}^{n}(z-\mu_j),
            \quad
            H_n(z)=h_0\prod_{l=1}^{n}(z-\nu_l).
         \end{equation}
Moreover, defining
      \begin{equation}\label{3.10}
        \hat{\mu}_j(x)
            =(\mu_j(x),-\mu_j(x)G_n(\mu_j(x),x))
            \in \mathcal{K}_{n}, ~
            j=0,\ldots,n,~x\in\mathbb{R},
      \end{equation}
and
       \begin{equation}\label{3.11}
        \hat{\nu}_l(x)
            =(\nu_l(x),\nu_l(x)G_n(\nu_l(x),x))
            \in \mathcal{K}_{n}, ~
            l=1,\ldots,n,~x\in\mathbb{R}.
      \end{equation}
Due to assumption (\ref{2.1}), $u$ is smooth and bounded, and hence
$F_{n+1}(z,x)$ and $H_n(z,x)$ share the same property. Thus, we
infers that
    \begin{equation}\label{3.12}
        \mu_j,\nu_l \in C(\mathbb{R}), \quad
        j=0,\dots,n,~l=1,\ldots,n.
    \end{equation}
here $\mu_j,\nu_l$ may have appropriate multiplicities.

The branch of $y(\cdot)$ near $P_{\infty_\pm}$ is fixed according to
       \begin{equation}\label{3.13}
         \underset{|z(P)| \rightarrow \infty \atop P \rightarrow P_{\infty_\pm}}{\mathrm{lim}}
         \frac{y(P)}{z(P) G_n(z(P),x)}=\mp 1.
       \end{equation}
Also by (\ref{3.7}), the divisor $(\phi(P,x))$ of $\phi(P,x)$ is
given by
         \begin{equation}\label{3.14}
           (\phi(P,x))=\mathcal{D}_{P_0\underline{\hat{\nu}}(x)}(P)
           -\mathcal{D}_{\hat{\mu}_0(x)\underline{\hat{\mu}}(x)}(P).
         \end{equation}
That means, $P_0,\hat{\nu}_1(x),\ldots,\hat{\nu}_{n}(x)$ are the
$n+1$ zeros of $\phi(P,x)$ and $\hat{\mu}_0(x),\hat{\mu}_1(x),
\ldots,$ $\hat{\mu}_{n}(x)$ are its $n+1$ poles. These zeros and
poles can be abbreviated in the following form
     \begin{equation}\label{3.15}
     \underline{\hat{\mu}}=\{\hat{\mu}_1,\ldots,\hat{\mu}_n\},
      \quad
      \underline{\hat{\nu}}=\{\hat{\nu}_1,\ldots,\hat{\nu}_n\}
      \in \mathrm{Sym}^n(\mathcal{K}_n).
     \end{equation}
Let us recall the holomorphic map (\ref{3.2}),
 \begin{eqnarray}\label{3.16}
       && \ast: \begin{cases}
                        \mathcal{K}_{n}\rightarrow\mathcal{K}_{n},
                       \\
                       P=(z,y_j(z))\rightarrow
                       P^\ast=(z,y_{j+1(\mathrm{mod}~
                       2)}(z)), \quad j=0,1,
                      \end {cases}
                     \nonumber \\
      && P^{\ast \ast}:=(P^\ast)^\ast, \quad \mathrm{etc}.,
       \end{eqnarray}
where $y_j(z),\, j=0,1$ satisfy $\mathcal{F}_{n}(z,y)=0$, namely
        \begin{equation}\label{3.17}
          (y-y_0(z))(y-y_1(z))
          =y^2-R_{2n+2}(z)=0.
        \end{equation}
Hence from (\ref{3.17}), we can easily get
        \begin{equation}\label{3.18}
           \begin{split}
             & y_0+y_1=0,\\
             & y_0y_1=-R_{2n+2}(z),\\
             & y_0^2+y_1^2=2R_{2n+2}(z).\\
           \end{split}
        \end{equation}
Further properties of $\phi(P,x)$ are summarized as follows.

\newtheorem{lem3.1}{Lemma}[section]
 \begin{lem3.1}
    Under the assumption $(\ref{2.1})$, let
    $P=(z,y)\in \mathcal{K}_{n}\setminus \{P_{\infty_+}, P_{\infty_-},P_0\},$ and
    $x \in \mathbb{R}$, and $u$ satisfies the $n$th
    stationary HS equation $(\ref{2.34})$.  Then
      \begin{equation}\label{3.19}
       \phi_x(P)+z^{-1}\phi(P)^2=-u_{xx},
      \end{equation}
       \begin{equation}\label{3.20}
        \phi(P)\phi(P^\ast)=-\frac{zH_n(z)}{F_{n+1}(z)},
       \end{equation}
       \begin{equation}\label{3.21}
        \phi(P)+\phi(P^\ast)=\frac{2zG_n(z)}{F_{n+1}(z)},
       \end{equation}
       \begin{equation}\label{3.22}
        \phi(P)-\phi(P^\ast)=\frac{2y}{F_{n+1}(z)}.
       \end{equation}
 \end{lem3.1}
 \textbf{Proof.}~~A direct calculation shows that (\ref{3.19}) holds.
Let us now prove (\ref{3.20})-(\ref{3.22}). Without loss of
generality, let $y_0(P)=y(P)$. From (\ref{3.7}), (\ref{2.19}) and
(\ref{3.18}), we arrive at
      \begin{eqnarray}\label{3.23}
      \phi(P)\phi(P^\ast) &=& \frac{y_0+zG_n}{F_{n+1}}
      ~\times ~ \frac{y_1+zG_n}{F_{n+1}}
        \nonumber \\
      &=& \frac{y_0y_1+(y_0+y_1)zG_n+z^2G_n^2}{F_{n+1}^2}
        \nonumber \\
      &=& \frac{-R_{2n+2}+z^2G_n^2}{F_{n+1}^2}
       = z\frac{-F_{n+1}H_n}{F_{n+1}^2}
         \nonumber \\
      &=& -\frac{zH_n}{F_{n+1}},
    \end{eqnarray}
    \begin{eqnarray}\label{3.24}
    \phi(P)+\phi(P^\ast) &=&  \frac{y_0+zG_n}{F_{n+1}}
      ~+ ~ \frac{y_1+zG_n}{F_{n+1}}
          \nonumber \\
      &=&  \frac{(y_0+y_1)+2zG_n}{F_{n+1}}
       = \frac{2zG_n}{F_{n+1}},
   \end{eqnarray}
   \begin{eqnarray}\label{3.25}
    \phi(P)-\phi(P^\ast) &=&  \frac{y_0+zG_n}{F_{n+1}}
      ~-~ \frac{y_1+zG_n}{F_{n+1}}
          \nonumber \\
      &=&  \frac{(y_0-y_1)}{F_{n+1}}
      = \frac{2y_0}{F_{n+1}}= \frac{2y}{F_{n+1}}.
   \end{eqnarray}
Hence we complete the proof. \quad $\square$
\\

Let us detail the properties of $\psi(P,x,x_0)$ below.

\newtheorem{lem3.2}[lem3.1]{Lemma}
 \begin{lem3.2}
    Under the assumption $(\ref{2.1})$,  let
    $P=(z,y)\in \mathcal{K}_{n}\setminus \{P_{\infty_+},P_{\infty_-},P_0\},$
    $(x,x_0)\in \mathbb{R}^2$, and  $u$ satisfies the $n$th
    stationary HS equation $(\ref{2.34})$. Then
      \begin{equation}\label{3.26}
        \psi_1(P,x,x_0)=\Big(\frac{F_{n+1}(z,x)}{F_{n+1}(z,x_0)}\Big)^{1/2}
         \mathrm{exp}\Big(\frac{y}{z}
          \int_{x_0}^x F_{n+1}(z,x^\prime)^{-1} dx^\prime
          \Big),
      \end{equation}
      \begin{equation}\label{3.27}
        \psi_1(P,x,x_0)\psi_1(P^\ast,x,x_0)=\frac{F_{n+1}(z,x)}{F_{n+1}(z,x_0)},
      \end{equation}
      \begin{equation}\label{3.28}
        \psi_2(P,x,x_0)\psi_2(P^\ast,x,x_0)=-\frac{H_n(z,x)}{z F_{n+1}(z,x_0)},
      \end{equation}
      \begin{equation}\label{3.29}
        \psi_1(P,x,x_0)\psi_2(P^\ast,x,x_0)
        +\psi_1(P^\ast,x,x_0)\psi_2(P,x,x_0)
        =2\frac{G_n(z,x)}{F_{n+1}(z,x_0)},
     \end{equation}
     \begin{equation}\label{3.30}
        \psi_1(P,x,x_0)\psi_2(P^\ast,x,x_0)
        -\psi_1(P^\ast,x,x_0)\psi_2(P,x,x_0)
        =\frac{-2y}{z F_{n+1}(z,x_0)}.
    \end{equation}
 \end{lem3.2}
 \textbf{Proof.}~~Equation (\ref{3.26}) can be proven through the following procedure.
Using (\ref{2.15}), the expression of $\psi_1$, (\ref{3.6}) and
(\ref{3.7}), we obtain
    \begin{eqnarray}\label{3.31}
        \psi_1(P,x,x_0) &=& \mathrm{exp}\left(z^{-1}\int_{x_0}^x
           \frac{y+zG_n(z,x^\prime)}{F_{n+1}(z,x^\prime)}~ dx^\prime
            \right)
           \\
        &=& \mathrm{exp}\left(z^{-1} \int_{x_0}^x
         \Big(   \frac{y}{F_{n+1}(z,x^\prime)}
         +\frac{1}{2}\frac{F_{n+1,x^\prime}(z,x^\prime)}{F_{n+1}(z,x^\prime)} \Big)
         dx^\prime \right), \nonumber
     \end{eqnarray}
which implies (\ref{3.26}). Moreover, (\ref{3.6}) and (\ref{3.8})
together with (\ref{3.20})-(\ref{3.22}) yields
     \begin{eqnarray}\label{3.32}
      \psi_1(P,x,x_0)\psi_1(P^\ast,x,x_0) &=&
       \mathrm{exp}\left(z^{-1}\int_{x_0}^x
          (\phi(P)+\phi(P^\ast))~ dx^\prime
         \right)
          \nonumber \\
       &=& \mathrm{exp}\left(z^{-1}\int_{x_0}^x
         \frac{2zG_n(z,x^\prime)}{F_{n+1}(z,x^\prime)} ~dx^\prime
         \right)
          \nonumber \\
       &=& \mathrm{exp}\left(\int_{x_0}^x
          \frac{F_{n+1,x^\prime}(z,x^\prime)}{F_{n+1}(z,x^\prime)} ~dx^\prime
         \right)
           \nonumber \\
       &=&
       \frac{F_{n+1}(z,x)}{F_{n+1}(z,x_0)},
    \end{eqnarray}
    \begin{eqnarray}\label{3.33}
       \psi_2(P,x,x_0)\psi_2(P^\ast,x,x_0) &=& z^{-2}
       \psi_1(P,x,x_0)\phi(P,x)\psi_1(P^\ast,x,x_0)\phi(P^\ast,x)
         \nonumber \\
       &=& z^{-2} \frac{F_{n+1}(z,x)}{F_{n+1}(z,x_0)}\frac{(-zH_n(z,x))}{F_{n+1}(z,x)}
          \nonumber \\
       &=& -\frac{H_n(z,x)}{zF_{n+1}(z,x_0)},
    \end{eqnarray}
    \begin{eqnarray}\label{3.34}
     &&
      \psi_1(P,x,x_0)\psi_2(P^\ast,x,x_0)+\psi_1(P^\ast,x,x_0)\psi_2(P,x,x_0)
      \nonumber \\
    && ~~~~~~~~
        =\psi_1(P)\psi_1(P^\ast)\phi(P^\ast)/z+
        \psi_1(P^\ast)\psi_1(P)\phi(P)/z
         \nonumber \\
    && ~~~~~~~~
       =\psi_1(P)\psi_1(P^\ast)(\phi(P)+\phi(P^\ast))/z
          \nonumber \\
    && ~~~~~~~~
       =\frac{F_{n+1}(z,x)}{F_{n+1}(z,x_0)}\frac{2zG_n(z,x)}{z F_{n+1}(z,x)}
         \nonumber \\
    && ~~~~~~~~
       =\frac{2G_n(z,x)}{F_{n+1}(z,x_0)},
    \end{eqnarray}
    \begin{eqnarray*}
     &&
      \psi_1(P,x,x_0)\psi_2(P^\ast,x,x_0)-\psi_1(P^\ast,x,x_0)\psi_2(P,x,x_0)
      \nonumber \\
    && ~~~~~~~~
        =\psi_1(P)\psi_1(P^\ast)\phi(P^\ast)/z+
        \psi_1(P^\ast)\psi_1(P)\phi(P)/z
         \nonumber \\
    && ~~~~~~~~
       =\psi_1(P)\psi_1(P^\ast)(\phi(P^\ast)-\phi(P))/z
       \end{eqnarray*}
    \begin{eqnarray}\label{3.35}
           &=&\frac{F_{n+1}(z,x)}{F_{n+1}(z,x_0)}\frac{-2y}{z F_{n+1}(z,x)}
         \nonumber \\
           &=&\frac{-2y}{z F_{n+1}(z,x_0)}.
    \end{eqnarray}
Hence (\ref{3.27})-(\ref{3.30}) hold. \quad $\square$

In Lemma 3.2 if we choose
      $$\psi_1(P)=\psi_{1,+}, ~ \psi_1(P^\ast)=\psi_{1,-},~
       \psi_2(P)=\psi_{2,+}, ~ \psi_2(P^\ast)=\psi_{2,-},$$
then (\ref{3.27})-(\ref{3.30}) imply
     \begin{equation}\label{3.36}
        (\psi_{1,+}\psi_{2,-}-\psi_{1,-}\psi_{2,+})^2=
          (\psi_{1,+}\psi_{2,-}+\psi_{1,-}\psi_{2,+})^2
          -4\psi_{1,+}\psi_{2,-}\psi_{1,-}\psi_{2,+},
     \end{equation}
which is equivalent to the basic identity (\ref{2.19}),
$z^2G_n^2+zF_{n+1}H_n=R_{2n+2}$. This fact reveals the relations
between our approach and the algebro-geometric solutions of the HS
hierarchy.

\newtheorem{rem3.3}[lem3.1]{Remark}
  \begin{rem3.3}
    The definition of stationary Baker-Akhiezer function $\psi$ of
    the HS hierarchy is analogous to that in the context of KdV or AKNS
    hierarchies. But the crucial difference is that $P_0$ is a
    essential singularity of $\psi$ in the HS hierarchy, which is the same as in
    CH hierarchy, but different from the KdV or AKNS
    hierarchy. This fact will be showed in the asymptotic expansions
    of $\psi$ in next section.
  \end{rem3.3}

Furthermore, we derive Dubrovin-type equations, which are
first-order coupled systems of differential equations and govern the
dynamics of the zeros $\mu_j(x)$ and $\nu_l(x)$ of $F_{n+1}(z,x)$
and $H_n(z,x)$ with respect to $x$. We recall the affine part of
$\mathcal{K}_n$ is nonsingular if
     \begin{equation}\label{3.37}
       \begin{split}
        & E_0=0,~~
        \{E_m\}_{m=1,\ldots,2n+1} \subset \mathbb{C} \setminus \{0\},
         \\
        & E_m \neq E_{m^\prime}
        \quad \textrm{for $m \neq m^\prime,
        m,m^\prime=1,\ldots,2n+1$}.
       \end{split}
     \end{equation}

\newtheorem{lem3.4}[lem3.1]{Lemma}
    \begin{lem3.4}
    Assume that $(\ref{2.1})$ holds and $u$ satisfies the $n$th
    stationary HS equation $(\ref{2.34})$.\\

        $(\mathrm{i})$
     If the zeros $\{\mu_j(x)\}_{j=0,\ldots,n}$
     of $F_{n+1}(z,x)$ remain distinct for $ x \in
     \Omega_\mu,$ where $\Omega_\mu \subseteq \mathbb{R}$ is an open
     interval, then
     $\{\mu_j(x)\}_{j=0,\ldots,n}$ satisfy the system of
     differential equations,
    \begin{equation}\label{3.38}
           \mu_{j,x}= 2 \frac{ y(\hat{\mu}_j)}{\mu_j}
            \prod_{\scriptstyle k=0 \atop \scriptstyle k \neq j }^{n}
            (\mu_j(x)-\mu_k(x))^{-1}, \quad j=0,\ldots,n,
        \end{equation}
   with initial conditions
       \begin{equation}\label{3.39}
         \{\hat{\mu}_j(x_0)\}_{j=0,\ldots,n}
         \in \mathcal{K}_{n},
       \end{equation}
   for some fixed $x_0 \in \Omega_\mu$. The initial value
   problems $(\ref{3.38})$, $(\ref{3.39})$ have a unique solution
   satisfying
        \begin{equation}\label{3.40}
         \hat{\mu}_j \in C^\infty(\Omega_\mu,\mathcal{K}_{n}),
         \quad j=0,\ldots,n.
        \end{equation}

    $(\mathrm{ii})$
     If the zeros $\{\nu_l(x)\}_{l=1,\ldots,n}$
     of $H_n(z,x)$ remain distinct for $ x \in
     \Omega_\nu,$ where $\Omega_\nu \subseteq \mathbb{R}$ is an open
     interval, then
     $\{\nu_l(x)\}_{l=1,\ldots,n}$ satisfy the system of
     differential equations,
    \begin{equation}\label{3.41}
           \nu_{l,x}=-2\frac{u_{xx}~y(\hat{\nu}_l)}{h_0 ~ \nu_l}
            \prod_{\scriptstyle k=1 \atop \scriptstyle k \neq l }^{n}
            (\nu_l(x)-\nu_k(x))^{-1}, \quad l=1,\ldots,n,
        \end{equation}
   with initial conditions
       \begin{equation}\label{3.42}
         \{\hat{\nu}_l(x_0)\}_{l=1,\ldots,n}
         \in \mathcal{K}_{n},
       \end{equation}
   for some fixed $x_0 \in \Omega_\nu$. The initial value
   problems $(\ref{3.41})$, $(\ref{3.42})$ have a unique solution
   satisfying
        \begin{equation}\label{3.43}
         \hat{\nu}_l \in C^\infty(\Omega_\nu,\mathcal{K}_{n}),
         \quad l=1,\ldots,n.
        \end{equation}
    \end{lem3.4}
\textbf{Proof.}~~For our convenience, let us focus on (\ref{3.38})
and (\ref{3.40}), the proof of (\ref{3.41}) and (\ref{3.43}) follows
in an identical manner. The derivatives of (\ref{3.9}) with respect
to $x$ take on
   \begin{equation}\label{3.44}
   F_{n+1,x}(\mu_j)= -\mu_{j,x}
   \prod_{\scriptstyle k=0 \atop \scriptstyle k \neq j }^{n}
   (\mu_j(x)-\mu_k(x)).
   \end{equation}
On the other hand, inserting $z=\mu_j$ into equation (2.15) leads to

       \begin{equation}\label{3.45}
        F_{n+1,x}(\mu_j)=2 G_n(\mu_j)
        = 2 \frac{y(\hat{\mu}_j)}{-\mu_j}.
       \end{equation}
Comparing (\ref{3.44}) with (\ref{3.45}) gives  ({\ref{3.38}). The
proof of smoothness assertion (\ref{3.40}) is analogous to
the mCH case in our latest paper \cite{18}. \quad $\square$\\

Let us now  turn to the trace formulas of the HS invariants, which
is the expressions of $f_{l}$ and $h_{l}$ in terms of symmetric
functions of the zeros $\mu_j$ and $\nu_l$ of $F_{n+1}$ and $H_n$,
respectively. Here, we just consider the simplest case.

\newtheorem{lem3.5}[lem3.1]{Lemma}
 \begin{lem3.5}
  If $(\ref{2.1})$ holds and $u$ satisfies the $n$th
  stationary HS equation $(\ref{2.34})$, then
    \begin{equation}\label{3.46}
    u= \frac{1}{2}\sum_{j=0}^n \mu_j -\frac{1}{2}\sum_{m=0}^{2n+1} E_m.
    \end{equation}
 \end{lem3.5}
\textbf{Proof.}~~By comparison of the coefficient of $z^{n}$
 of $F_{n+1}$ in (\ref{2.28}) and (\ref{3.9}),
 taking account into (\ref{2.6}) yields
      \begin{equation}\label{3.47}
        -2u+c_1=-\sum_{j=0}^n \mu_j.
      \end{equation}
 The constant $c_1$ can be determined by a long straightforward calculation
 comparing the coefficients of $z^{2n+1}$ in (\ref{2.19}), which leads to
     \begin{equation}\label{3.48}
        c_1=-\sum_{m=0}^{2n+1} E_m.
     \end{equation}

 \section{Stationary algebro-geometric solutions of HS hierarchy}
 In this section we continue our study of the stationary HS
 hierarchy, and will obtain explicit Riemann theta function
 representations for the meromorphic function $\phi$, and especially,
 for the potentials $u$ of the stationary HS hierarchy.

Let us begin with the asymptotic properties of $\phi$ and
$\psi_j,j=1,2$.

\newtheorem{lem4.1}{Lemma}[section]
   \begin{lem4.1}
    Assume that $(\ref{2.1})$ to hold and $u$ satisfies the $n$th
    stationary HS equation $(\ref{2.34})$. Moreover, let $P=(z,y)
    \in \mathcal{K}_n \setminus \{P_{\infty_+}, P_{\infty_-}, P_0\},$ $(x,x_0) \in
    \mathbb{R}^2$. Then
     \begin{equation}\label{4.1}
        \phi(P) \underset{\zeta \rightarrow 0}{=} -u_x+O(\zeta),
        \qquad P \rightarrow P_{\infty_\pm}, \quad \zeta=z^{-1},
     \end{equation}
     \begin{equation}\label{4.2}
         \phi(P) \underset{\zeta \rightarrow 0}{=}
          i~a \Big( \prod_{m=1}^{2n+1} E_m \Big)^{1/2} f_{n+1}^{-1} \zeta
          +O(\zeta^2),
           \qquad P \rightarrow P_{0}, \quad \zeta=z^{1/2},
     \end{equation}
and
     \begin{eqnarray}
     \psi_1(P,x,x_0) &\underset{\zeta \rightarrow 0}{=}&
       \mathrm{exp} \Big(
       (u(x_0)-u(x))\zeta+ O(\zeta^2) \Big),
        \\
       &&~~~~~
        P \rightarrow P_{\infty_\pm},~ \zeta=z^{-1},
          \nonumber \\
       \psi_2(P,x,x_0) &\underset{\zeta \rightarrow 0}{=}&
        O(\zeta)~
        \mathrm{exp} \Big(
        (u(x_0)-u(x))\zeta+ O(\zeta^2) \Big),
          \\
        &&~~~~~
         P \rightarrow P_{\infty_\pm},~ \zeta=z^{-1},
          \nonumber
     \end{eqnarray}
      \begin{eqnarray}
         \psi_1(P,x,x_0) &\underset{\zeta \rightarrow 0}{=}&
         \mathrm{exp} \Big( \frac{i}{\zeta} \int_{x_0}^x dx^\prime
          ~a \Big( \prod_{m=1}^{2n+1} E_m \Big)^{1/2} f_{n+1}(x^\prime)^{-1}
          +O(1) \Big), \nonumber \\
          && ~~~~~
           P \rightarrow P_{0}, \quad \zeta=z^{1/2},
            \\
         \psi_2(P,x,x_0) &\underset{\zeta \rightarrow 0}{=}&
         O(\zeta^{-1})~
        \mathrm{exp} \Big( \frac{i}{\zeta} \int_{x_0}^x dx^\prime
          ~a \Big( \prod_{m=1}^{2n+1} E_m \Big)^{1/2} f_{n+1}(x^\prime)^{-1}
          +O(1) \Big), \nonumber \\
          && ~~~~~
           P \rightarrow P_{0}, \quad \zeta=z^{1/2}.
      \end{eqnarray}
  \end{lem4.1}
  \textbf{Proof.}~~Under the local coordinates $\zeta=z^{-1}$
near $P_{\infty_\pm}$ and $\zeta=z^{1/2}$ near $P_0$, the existence of the
asymptotic expansions of $\phi$ is clear from its explicit
expressions in (\ref{3.7}). Next, we use the Riccati-type equation
(\ref{3.19}) to compute the explicit expansion coefficients.
Inserting the ansatz
  \begin{equation}\label{4.7}
     \phi \underset{z \rightarrow \infty}{=}
     \phi_0 + \phi_1 z^{-1}+O(z^{-2})
  \end{equation}
into (\ref{3.19}) and comparing the powers of $z^0$ then yields
(\ref{4.1}). Similarly, inserting the ansatz
    \begin{equation}\label{4.8}
     \phi \underset{z \rightarrow 0}{=}
       \phi_{1} z^{1/2} + \phi_2 z  +O(z^{3/2})
    \end{equation}
into (\ref{3.19}) and comparing the power of $z^0$ then yields
(\ref{4.2}), where we used (\ref{2.31}) and
     \begin{equation}\label{4.9}
        f_{n+1}h_n=- a^2 \prod_{m=1}^{2n+1}E_m,
     \end{equation}
which can be obtained by (\ref{2.19}). Finally, expansions
(4.3)-(4.6) follow up by (\ref{3.6}), (\ref{3.8}), (\ref{4.1}) and
(\ref{4.2}). \qquad $\square$

\newtheorem{rem4.2}[lem4.1]{Remark}
  \begin{rem4.2}
    From $(4.5)$ and $(4.6)$, we note the unusual fact
    that $P_0$ is the essential singularity
    of $\psi_j$, $j=1,2$, this is consistent with Remark $3.3$. Also the
    leading-order exponential term $\psi_j$, $j=1,2,$ near $P_0$ is
    $x$-dependent, which makes matters worse.
    This is in sharp contrast to standard
    Baker-Akhiezer functions that typically feature a linear
    behavior with respect to $x$ such as
    $\mathrm{exp}(c(x-x_0)\zeta^{-1})$
    near $P_0$.
  \end{rem4.2}

Let us now introduce the holomorphic differentials $\eta_l(P)$ on
$\mathcal{K}_{n}$
      \begin{equation}\label{4.10}
        \eta_l(P)= \frac{a~z^{l-1}}{y(P)} dz,
        \qquad l=1,\ldots,n,
      \end{equation}
and choose a homology basis $\{a_j,b_j\}_{j=1}^{n}$ on
$\mathcal{K}_{n}$ in such a way that the intersection matrix of the
cycles satisfies
$$a_j \circ b_k =\delta_{j,k},\quad a_j \circ a_k=0, \quad
b_j \circ   b_k=0, \quad j,k=1,\ldots, n.$$ Define an invertible
matrix $E \in GL(n, \mathbb{C})$ as
follows
   \begin{equation}\label{4.11}
          \begin{split}
        & E=(E_{j,k})_{n \times n}, \quad E_{j,k}=
           \int_{a_k} \eta_j, \\
        &  \underline{c}(k)=(c_1(k),\ldots, c_{n}(k)), \quad
           c_j(k)=(E^{-1})_{j,k},
           \end{split}
   \end{equation}
and the normalized holomorphic differentials
        \begin{equation}\label{4.12}
          \omega_j= \sum_{l=1}^{n} c_j(l)\eta_l, \quad
          \int_{a_k} \omega_j = \delta_{j,k}, \quad
          \int_{b_k} \omega_j= \tau_{j,k}, \quad
          j,k=1, \ldots ,n.
        \end{equation}
Apparently, the matrix $\tau$ is symmetric and has a
positive-definite imaginary part.

The symmetric function $\Phi_{n}^{(j)}(\bar{\mu})$ and
$\Psi_{n+1}(\bar{\mu})$ are defined by
     \begin{equation}\label{4.13}
        \Phi_{n}^{(j)}(\bar{\mu})=(-1)^{n}
          \prod_{\scriptstyle p=0 \atop \scriptstyle p \neq j}^n \mu_p,
     \end{equation}
     \begin{equation}\label{4.14}
        \Psi_{n+1}(\bar{\mu})=(-1)^{n+1} \prod_{p=0}^n \mu_p.
     \end{equation}

The following result shows that the
nonlinearity of the Abel map in the HS hierarchy. This feature is
analogous to CH hierarchy but sharp apposed to other integrable
soliton equations such as KdV and AKNS hierarchies.

\newtheorem{the4.3}[lem4.1]{Theorem}
  \begin{the4.3}
    Assume $(\ref{2.1})$ to hold and
    suppose that
    $\{\hat{\mu}_j\}_{j=0,\ldots,n}$ satisfies the stationary
    Dubrovin equations $(\ref{3.38})$ on $\Omega_\mu$ and
    remain distinct for $ x \in
     \Omega_\mu,$ where $\Omega_\mu \subseteq \mathbb{R}$ is an open
     interval. Introducing the associated divisor
     $\mathcal{D}_{\hat{\mu}_0(x) \underline{\hat{\mu}}(x)}$. Then
      \begin{equation}\label{4.15}
            \partial_x
            \underline{\alpha}_{Q_0}( \mathcal{D}_{\hat{\mu}_0(x) \underline{\hat{\mu}}(x)})
            = -\frac{2a}{ \Psi_{n+1}(\bar{\mu}(x))}
             \underline{c}(1),
            \qquad    x \in \Omega_\mu.
        \end{equation}
   In particular, the Abel map does not linearize the divisor
   $\mathcal{D}_{\hat{\mu}_0(x)\underline{\hat{\mu}}(x)}$ on $\Omega_\mu$,
   where $\bar{\mu}(x)=(\mu_0(x),\mu_1(x),\ldots,
   \mu_n(x))=\mu_0(x) \underline{\mu}(x).$
  \end{the4.3}
\textbf{Proof.}~~Is easy to see that
        \begin{equation}\label{4.16}
 \frac{1}{\mu_j}=
        \frac{\prod_{\scriptstyle p=0 \atop \scriptstyle p \neq j}^n \mu_p}
            {\prod_{p=0}^n \mu_p}
        =-\frac{
        \Phi_{n}^{(j)}(\bar{\mu})}{\Psi_{n+1}(\bar{\mu})},
        \quad j=1,\ldots,n.
        \end{equation}
Let
        \begin{equation}\label{4.17}
        \underline{\omega}=(\omega_1,\ldots,\omega_n),
        \end{equation}
and choose a appropriate base point $Q_0$. Then we arrive at
     \begin{eqnarray*}
      &&
      \partial_x  \underline{\alpha}_{Q_0}(\mathcal{D}_{\hat{\mu}_0(x)\underline{\hat{\mu}}(x)})
 =
 \partial_x  \Big(\sum_{j=0}^n \int_{Q_0}^{\hat{\mu}_j} \underline{\omega} \Big)
 =\sum_{j=0}^n \mu_{j,x} \sum_{k=1}^n \underline{c}(k)
  \frac{a~\mu_j^{k-1}}{y(\hat{\mu}_j)}
      \nonumber \\
   &&=
    \sum_{j=0}^n \sum_{k=1}^n
    \frac{2a~\mu_j^{k-1}}{\mu_j}
    \frac{1}
    {\prod_{\scriptstyle l=0 \atop \scriptstyle l \neq j }^{n}(\mu_j-\mu_l)}
    \underline{c}(k)
     \nonumber \\
    &&
    =
     -\frac{2a}{\Psi_{n+1}(\bar{\mu})}
     \sum_{j=0}^n \sum_{k=1}^n \underline{c}(k)
       \frac{\mu_j^{k-1}}
    {\prod_{\scriptstyle l=0 \atop \scriptstyle l \neq j }^{n}(\mu_j-\mu_l)}
     \Phi_{n}^{(j)}(\bar{\mu})
       \nonumber \\
    \end{eqnarray*}
    \begin{eqnarray}\label{4.18}
    &&
     =
     -\frac{2a}{\Psi_{n+1}(\bar{\mu})}
     \sum_{j=0}^n \sum_{k=1}^n \underline{c}(k)
     (U_{n+1}(\bar{\mu}))_{k,j}(U_{n+1}(\bar{\mu}))_{j,1}^{-1}
       \nonumber \\
    &&
    =
     -\frac{2a}{\Psi_{n+1}(\bar{\mu})}
      \sum_{k=1}^n \underline{c}(k) \delta_{k,1}
      \nonumber \\
    &&
    =
      -\frac{2a}{\Psi_{n+1}(\bar{\mu})}
        \underline{c}(1),
    \end{eqnarray}
where we used
    \begin{equation}\label{4.19}
      (U_{n+1}(\bar{\mu}))=
       \Big( \frac{\mu_j^{k-1}}
    {\prod_{\scriptstyle l=0 \atop \scriptstyle l \neq j }^{n}(\mu_j-\mu_l)}
    \Big)_{\scriptstyle j=0 \atop \scriptstyle k=1}^n,
    \quad
     (U_{n+1}(\bar{\mu}))^{-1}=
     \Big(\Phi_{n}^{(j)}(\bar{\mu})\Big)_{j=0}^n,
    \end{equation}
the definition of which is analogous to (E.25) and (E.26) in
\cite{15}.
\quad $\square$ \\

The analogous results hold for the corresponding divisor
$\mathcal{D}_{\underline{\hat{\nu}}(x)}$ associated with
$\phi(P,x)$ can be obtained in the same
way.\\

Next, we introduce \footnote{ Here we choose the same path of integration
from $Q_0$ and $P$ in all integrals in (\ref{4.20}) \\ and
(\ref{4.21}).}
   \begin{equation}\label{4.20}
     \begin{split}
    & \underline{\widehat{B}}_{Q_0}: \mathcal{K}_n \setminus
    \{P_{\infty_+}, P_{\infty_-}\} \rightarrow \mathbb{C}^n, \\
    & P \mapsto \underline{\widehat{B}}_{Q_0}(P)
    =(\widehat{B}_{Q_0,1},\ldots,\widehat{B}_{Q_0,n}) \\
    & ~~~~~~~~~~~~~~~~~
     =
      \begin{cases}
       \int_{Q_0}^P \tilde{\omega}_{P_{\infty_+}, P_{\infty_-}}^{(3)},
         ~~~~~~~~~~~~~n=1 \\
       \Big(\int_{Q_0}^P \eta_2,\ldots, \int_{Q_0}^P \eta_n,
           \int_{Q_0}^P \tilde{\omega}_{P_{\infty_+}, P_{\infty_-}}^{(3)}
       \Big),~~~~~~~~~~~ n \geq 2,
      \end{cases}
     \end{split}
   \end{equation}
where
$$\tilde{\omega}_{P_{\infty_+}, P_{\infty_-}}^{(3)}=
\frac{a~z^{n}}{y(P)}dz$$
denotes a differential of the third kind with simple poles at
$P_{\infty_+}$ and $P_{\infty_-}$ and corresponding residues
$+1$ and $-1$, respectively.
Moreover,
   \begin{equation}\label{4.21}
     \begin{split}
       & \underline{\hat{\beta}}_{Q_0}:
         \mathrm{Sym}^n(
       \mathcal{K}_n \setminus
    \{P_{\infty_+}, P_{\infty_-}\}) \rightarrow \mathbb{C}^n, \\
      & \mathcal{D}_{\underline{Q}} \mapsto
      \underline{\hat{\beta}}_{Q_0}(\mathcal{D}_{\underline{Q}})
      =\sum_{j=1}^n \underline{\widehat{B}}_{Q_0}(Q_j), \\
      &
      \underline{Q}=\{Q_1,\ldots,Q_n\}
      \in \mathrm{Sym}^n( \mathcal{K}_n \setminus
    \{P_{\infty_+}, P_{\infty_-}\}).
     \end{split}
   \end{equation}

The following result is a special case of Theorem 4.3, which will be used
to provide the proper change of variables to linear the divisor
$\mathcal{D}_{\hat{\mu}_0(x) \underline{\hat{\mu}}(x)}$ associated with
$\phi(P,x)$.

\newtheorem{the4.4}[lem4.1]{Theorem}
  \begin{the4.4}
    Assume that $(\ref{2.1})$ holds and
    the statements of $\mu_j$ in Theorem $4.3$ are all true. Then
    \begin{equation}\label{4.22}
        \partial_x \sum_{j=0}^n \int_{Q_0}^{\hat{\mu}_j(x)} \eta_1
        = -\frac{2a}{\Psi_{n+1}(\bar{\mu}(x))},
        \qquad x\in \Omega_\mu,
    \end{equation}
    \begin{equation}\label{4.23}
        \partial_x \underline{\hat{\beta}}
        (\mathcal{D}_{ \underline{\hat{\mu}}(x)})
        =
        \begin{cases}
         2a,
         ~~~~~~~~n=1,\\
          2a~(0,\ldots,0,1),
          ~~~~~~~~n\geq 2,
        \end{cases}
        \quad  x\in \Omega_\mu.
    \end{equation}
  \end{the4.4}
\textbf{Proof.}~~Equations (\ref{4.22}) is a special case
(\ref{4.15}) and (\ref{4.23}) follows from (\ref{4.18}).
Alternatively, one can follow the same way as shown in Theorem 4.3
to derive
(\ref{4.22}) and (\ref{4.23}). \quad   $\square$ \\

Let $\theta(\underline{z})$ denote the Riemann theta function
associated with $\mathcal{K}_{n}$ and an appropriately fixed
homology basis. We assume $\mathcal{K}_{n}$ to be nonsingular. Next,
choosing a convenient base point $Q_0 \in \mathcal{K}_{n} \setminus
\{\hat{\mu}_0(x),P_0\}$, the vector of Riemann constants
$\underline{\Xi}_{Q_0}$ is given by (A.66) \cite{15}, and the Abel
maps $\underline{A}_{Q_0}(\cdot) $ and
$\underline{\alpha}_{Q_0}(\cdot)$ are defined by

\begin{equation}\label{4.24}
    \begin{split}
    &
   \underline{A}_{Q_0}:\mathcal{K}_{n} \rightarrow
   J(\mathcal{K}_{n})=\mathbb{C}^{n}/L_{n}, \\
   &
   P \mapsto \underline{A}_{Q_0} (P)=(\underline{A}_{Q_0,1}(P),\ldots,
  \underline{A}_{Q_0,n} (P))
    =\left(\int_{Q_0}^P\omega_1,\ldots,\int_{Q_0}^P\omega_{n}\right)
  (\mathrm{mod}~L_{n})
  \end{split}
  \end{equation}
   and
   \begin{equation}\label{4.25}
     \begin{split}
   & \underline{\alpha}_{Q_0}:
   \mathrm{Div}(\mathcal{K}_{n}) \rightarrow
   J(\mathcal{K}_{n}),\\
   & \mathcal{D} \mapsto \underline{\alpha}_{Q_0}
   (\mathcal{D})= \sum_{P\in \mathcal{K}_{n}}
    \mathcal{D}(P)\underline{A}_{Q_0} (P),
    \end{split}
    \end{equation}
where $L_{n}=\{\underline{z}\in \mathbb{C}^{n}|
           ~\underline{z}=\underline{N}+\tau\underline{M},
           ~\underline{N},~\underline{M}\in \mathbb{Z}^{n}\}.$ \\

Let
   \begin{equation}\label{4.26}
    \omega_{\hat{\mu}_0(x) P_0}^{(3)}(P)=
     \frac{a}{y} \prod_{j=1}^n(z-\lambda_j) dz
    \end{equation}
be the normalized differential of the third kind  holomorphic on
$\mathcal{K}_{n} \setminus \{\hat{\mu}_0(x), P_0\}$ with simple poles at
$\hat{\mu}_0(x)$ and $P_0$ with residues $\pm 1$, respectively, that is,
 \begin{equation}\label{4.27}
            \begin{split}
              & \omega_{\hat{\mu}_0(x) P_0}^{(3)}(P) \underset
              {\zeta \rightarrow 0}{=} (\zeta^{-1}+O(1))d \zeta,
              \quad \textrm{as $P \rightarrow \hat{\mu}_0(x),$}\\
              & \omega_{\hat{\mu}_0(x) P_0}^{(3)}(P) \underset
              {\zeta \rightarrow 0}{=} (-\zeta^{-1}+O(1))d \zeta,
              \quad \textrm{as $P \rightarrow P_0,$}
            \end{split}
         \end{equation}
where the local coordinate are given by
    \begin{equation}\label{4.28}
        \zeta=z^{-1}
        ~~ \textrm{for $P$ near $ \hat{\mu}_0(x)$},
        \qquad
        \zeta= z^{1/2}
        ~~\textrm{for $P$ near $P_0$},
    \end{equation}
and the constants $\{\lambda_j\}_{j=1,\ldots,n}$ are determined by
the normalization condition
      $$\int_{a_k} \omega_{\hat{\mu}_0(x) P_0}^{(3)}=0,
      \qquad k=1,\ldots,n.$$
Then
 \begin{equation} \label{4.29}
   \int_{Q_0}^P \omega_{\hat{\mu}_0(x) P_0}^{(3)}(P) \underset
  {\zeta \rightarrow 0}{=} \mathrm{ln} \zeta +
  e_0+O(\zeta),
  \quad \textrm{as $P \rightarrow \hat{\mu}_0(x),$}
  \end{equation}
  \begin{equation}\label{4.30}
    \int_{Q_0}^P \omega_{\hat{\mu}_0(x) P_0}^{(3)}(P) \underset
    {\zeta \rightarrow 0}{=} -\mathrm{ln} \zeta +
    d_0+O(\zeta),
     \quad \textrm{as $P \rightarrow P_0,$}
    \end{equation}
for some constants $e_0,d_0 \in \mathbb{C}$ that arise from the
integrals at their lower limits $Q_0$. We also note that
  \begin{equation}\label{4.31}
   \underline{A}_{Q_0}(P)-\underline{A}_{Q_0}(P_{\infty_\pm})
   \underset {\zeta \rightarrow 0}{=}
   \pm \underline{U}\zeta+O(\zeta^2),
   \quad \textrm{as $P \rightarrow P_{\infty_\pm},$}
   \quad \underline{U}=\underline{c}(n).
  \end{equation}
The following abbreviations are used for our convenience:
    \begin{eqnarray}\label{4.32}
         &&
           \underline{z}(P,\underline{Q})= \underline{\Xi}_{Q_0}
           -\underline{A}_{Q_0}(P)+\underline{\alpha}_{Q_0}
             (\mathcal{D}_{\underline{Q}}), \nonumber \\
          &&
           P\in \mathcal{K}_{n},\,
           \underline{Q}=(Q_1,\ldots,Q_{n})\in
           \mathrm{Sym}^{n}(\mathcal{K}_{n}),
         \end{eqnarray}
where $\underline{z}(\cdot,\underline{Q}) $ is independent of the
choice of base point $Q_0$. \\

Moreover, from Theorem 4.3 and Theorem 4.4 we note that the Abel map
dose not linearize the divisor $\mathcal{D}_{\hat{\mu}_0(x)
\underline{\hat{\mu}}(x)}$. However, the change of variables
     \begin{equation}\label{4.33}
        x \mapsto \tilde{x}= \int^x dx^\prime
        \Big( \frac{2a}{\Psi_{n+1}(\bar{\mu}(x^\prime))}
                \Big)
     \end{equation}
linearizes the Abel map
$\underline{A}_{Q_0}(\mathcal{D}_{\hat{\tilde{\mu}}_0(\tilde{x})
\underline{\hat{\tilde{\mu}}}(\tilde{x})}),$
$\tilde{\mu}_j(\tilde{x})=\mu_j(x),j=0,\ldots,n.$ The intricate
relation between the variable $x$ and $\tilde{x}$  is discussed
detailedly in Theorem 4.5. \\

Based on the above all these preparations, let us  now give an
explicit representations for the meromorphic function $\phi$ and the
stationary HS solutions $u$ in terms of the Riemann theta function
associated with $\mathcal{K}_{n}$. Here we assume the affine part of
$\mathcal{K}_{n}$ to be nonsingular.

\newtheorem{the4.5}[lem4.1]{Theorem}
 \begin{the4.5}
 Assume that the curve $\mathcal{K}_{n}$ is nonsingular,
 $(\ref{2.1})$ holds and $u$ satisfies the $n$th
stationary HS equation $(\ref{2.34})$ on $\Omega$. Moreover, let
$P=(z,y) \in \mathcal{K}_n \setminus \{P_0\},$ and $x \in \Omega$,
where $ \Omega \subseteq \mathbb{R}$ is an open interval. In
addition, suppose that $\mathcal{D}_{\underline{\hat{\mu}}(x)}$, or
equivalently $\mathcal{D}_{\underline{\hat{\nu}}(x)}$ is nonspecial
for $x\in \Omega$. Then, $\phi$ and $u$ have the following
representations
    \begin{eqnarray}\label{4.34}
    \phi(P,x)&=&i a \Big(\prod_{m=1}^{2n+1} E_m \Big)^{1/2} f_{n+1}^{-1}
    \frac{\theta(\underline{z}(P,\underline{\hat{\nu}}(x)))
            \theta(\underline{z}(P_{0},\underline{\hat{\mu}}(x)))}
            {\theta(\underline{z}(P_{0},\underline{\hat{\nu}}(x)))
            \theta(\underline{z}(P,\underline{\hat{\mu}}(x)))}
             \nonumber \\
    && \times~
            \mathrm{exp}\left(d_0
            -\int_{Q_0}^P \omega_{\hat{\mu}_0(x) P_0}^{(3)}\right),
  \end{eqnarray}
   \begin{eqnarray}\label{4.35}
    u(x)&=&-\frac{1}{2} \sum_{m=0}^{2n+1} E_m +\frac{1}{2} \sum_{j=1}^n \lambda_j
     \nonumber \\
    &&
    -\frac{1}{2} \sum_{j=1}^nU_j \partial_{\omega_j} \mathrm{ln}
\left(\frac{\theta(\underline{z}(P_{\infty_+},\underline{\hat{\mu}}(x))+\underline{\omega})}
 {\theta(\underline{z}(P_{\infty_-},\underline{\hat{\mu}}(x))+\underline{\omega})}\right)
     \Big|_{\underline{\omega}=0}.
   \end{eqnarray}
Moreover, let $\mu_j,$ $j=0,\ldots,n$ be  not vanishing on $\Omega$
and $x,x_0 \in \Omega.$ Then, we have the following constraint
  \begin{eqnarray}\label{4.36}
        2a(x-x_0)
        &=&-2a \int_{x_0}^x
        \frac{dx^\prime}{\prod_{k=0}^n \mu_k(x^\prime)}
        \sum_{j=1}^n
        \Big(\int_{a_j} \tilde{\omega}_{P_{\infty_+} P_{\infty_-}}^{(3)} \Big)
        c_j(1) \nonumber \\
        &+&
         \mathrm{ln}
        \left(
        \frac{\theta(\underline{z}(P_{\infty_-},\underline{\hat{\mu}}(x_0)))
            \theta(\underline{z}(P_{\infty_+},\underline{\hat{\mu}}(x)))}
            {\theta(\underline{z}(P_{\infty_+},\underline{\hat{\mu}}(x_0)))
            \theta(\underline{z}(P_{\infty_-},\underline{\hat{\mu}}(x)))}
        \right)
  \end{eqnarray}
and
   \begin{eqnarray}\label{4.37}
      \underline{\hat{\alpha}}_{Q_0}( \mathcal{D}_{\hat{\mu}_0(x) \underline{\hat{\mu}}(x)})
      &=&
      \underline{\hat{\alpha}}_{Q_0}(\mathcal{D}_{\hat{\mu}_0(x_0) \underline{\hat{\mu}}(x_0)})
       -2a \int_{x_0}^x
       \frac{dx^\prime}{ \Psi_{n+1}(\bar{\mu}(x^\prime))}
             \underline{c}(1)
          \nonumber \\
       &=&
       \underline{\hat{\alpha}}_{Q_0}(\mathcal{D}_{\hat{\mu}_0(x_0) \underline{\hat{\mu}}(x_0)})
       - \underline{c}(1)(\tilde{x}-\tilde{x}_0).
     \end{eqnarray}
\end{the4.5}
\textbf{Proof.}~~First, let us assume
    \begin{equation}\label{4.38}
      \mu_j(x)\neq \mu_{j^\prime}(x), \quad \nu_k(x)\neq \nu_{k^\prime}(x)
      \quad \textrm{for $j\neq j^\prime, k\neq k^\prime$ and
      $x\in\widetilde{\Omega}$},
      \end{equation}
 where $\widetilde{\Omega}\subseteq\Omega$.
 From (\ref{3.14}), $\mathcal
 {D}_{P_{0}\underline{\hat{\nu}}}\sim
 \mathcal {D}_{\hat{\mu}_0 \underline{\hat{\mu}}}$, and
 $(P_0)^\ast \notin\{\hat{\nu}_1,\cdots,\hat{\nu}_n \}$
 by hypothesis, one can use Theorem A.31 \cite{15} to
 conclude that $\mathcal {D}_{\underline{\hat{\mu}}}
 \in \textrm{Sym}^n(\mathcal {K}_n)$ is nonspecial. This
 argument is of course symmetric with respect to
 $\underline{\hat{\mu}}$ and $\underline{\hat{\nu}}$. Thus, $\mathcal
 {D}_{\underline{\hat{\mu}}}$ is nonspecial if and only
 if $\mathcal{D}_{\underline{\hat{\nu}}}$ is.

 Next, we derive the
 representations of $\phi$ and $u$ in terms of the Riemann theta
 function. A special case of Riemann's vanishing theorem (Theorem A.26 \cite{15})
 yields
  \begin{equation}\label{4.39}
        \theta(\underline{\Xi}_{Q_0}-\underline{A}_{Q_0}(P)+\underline{\alpha}_{Q_0}(\mathcal
        {D}_{\underline{Q}}))=0 \quad \textrm{ if and only if $P\in
        \{Q_1,\cdots,Q_n\}$}.
  \end{equation}
Therefore, the divisor (\ref{3.14}) of $\phi(P,x)$ suggests
considering expressions of the following type
\begin{equation}\label{4.40}
C(x)\frac{
\theta(\underline{\Xi}_{Q_0}-\underline{A}_{Q_0}(P)+\underline{\alpha}_{Q_0}(\mathcal
{D}_{\underline{\hat{\nu}}(x)}))}{
\theta(\underline{\Xi}_{Q_0}-\underline{A}_{Q_0}(P)+\underline{\alpha}_{Q_0}(\mathcal
{D}_{\underline{\hat{\mu}}(x)}))} \mathrm{exp}\Big(d_0-\int_{Q_0}^P
\omega_{\hat{\mu}_0(x) P_{0}}^{(3)}\Big),
\end{equation}
where $C(x)$ is independent of $P\in\mathcal {K}_n$. So, together
with the asymptotic expansion of $\phi(P,x)$ near $P_0$ in
(\ref{4.2}), we are able to obtain (\ref{4.34}). The representation
(\ref{4.35}) for $u$ on $\widetilde{\Omega}$ follows from trace
formula (\ref{3.46}) and the expression (F.88 \cite{15}) for
$\sum_{j=0}^n \mu_j$.

To prove the constraint (\ref{4.36}), one can refs Theorem 4.5 in
our latest paper \cite{18}. Equations (\ref{4.37}) is clear from
(\ref{4.15}). Finally, the extension of all results from $x \in
\widetilde{\Omega}$ to $x \in   \Omega$  follows by the continuity
of $\underline{\alpha}_{Q_0}$ and the hypothesis of
$\mathcal{D}_{\underline{\hat{\mu}}(x)}$ being nonspecial for $x \in
\Omega$. \quad $\square$

\newtheorem{rem4.6}[lem4.1]{Remark}
  \begin{rem4.6}
  The stationary HS solutions $u$ in $(\ref{4.35})$ is a
  quasi-periodic function with respect to the new variable
  $\tilde{x}$ in $(\ref{4.33})$. The Abel map in $(\ref{4.37})$
  linearize the divisor
  $\mathcal{D}_{\hat{\mu}_0(x) \underline{\hat{\mu}}(x)}$ on $\Omega$ with respect
  to $\tilde{x}$.
  \end{rem4.6}

  \newtheorem{rem4.7}[lem4.1]{Remark}
  \begin{rem4.7}
   The similar results to $(\ref{4.36})$ and $(\ref{4.37})$ (i.e.
   the Abel map also linearize the divisor
   $\mathcal{D}_{\underline{\hat{\nu}}(x)}$ on $\Omega$ with respect
   to $\bar{x}$) hold for the divisor $\mathcal{D}_{\underline{\hat{\nu}}(x)}$
   associated with $\phi(P,x)$. The change of variables is
   \begin{equation}\label{4.41}
        x \mapsto \bar{x}= \int^x dx^\prime
        \Big( \frac{1}{\Psi_n(\underline{\nu}(x^\prime))}
         \frac{u_{x^\prime x^\prime}}{h_0(x^\prime)}
        \Big).
     \end{equation}
  \end{rem4.7}

  \newtheorem{rem4.8}[lem4.1]{Remark}
  \begin{rem4.8}
    Since  $\mathcal
 {D}_{P_{0}\underline{\hat{\nu}}}$ and
 $\mathcal {D}_{\hat{\mu}_0 \underline{\hat{\mu}}}$
 are linearly equivalent, that is
    \begin{equation}\label{4.42}
    \underline{A}_{Q_0}(\hat{\mu}_0(x)) + \underline{\alpha}_{Q_0}
    (\mathcal{D}_{\underline{\hat{\mu}}(x)})
    =\underline{A}_{Q_0}(P_0)+ \underline{\alpha}_{Q_0}
    (\mathcal{D}_{\underline{\hat{\nu}}(x)}).
    \end{equation}
Then we infer
    \begin{equation}\label{4.43}
        \underline{\alpha}_{Q_0}
    (\mathcal{D}_{\underline{\hat{\nu}}(x)})
    =\underline{\Delta}+ \underline{\alpha}_{Q_0}
    (\mathcal{D}_{\underline{\hat{\mu}}(x)}),
    \qquad \underline{\Delta}=\underline{A}_{P_0}(\hat{\mu}_0(x)).
    \end{equation}
Hence
   \begin{equation}\label{4.44}
    \underline{z}(P,\underline{\hat{\nu}})=
    \underline{z}(P,\underline{\hat{\mu}})+\underline{\Delta},
    \qquad P\in \mathcal{K}_n.
   \end{equation}
The representations of $\phi$ and $u$ in $(\ref{4.34})$ and
$(\ref{4.35})$ can be rewritten in terms of
$\mathcal{D}_{\underline{\hat{\nu}}(x)}$ respectively.
  \end{rem4.8}

\newtheorem{rem4.9}[lem4.1]{Remark}
  \begin{rem4.9}
   We have emphasized in Remark $4.2$ that the Baker-Akhiezer
   functions $\psi$ in $(\ref{3.6})$ and $(\ref{3.8})$ for the  HS hierarchy
   enjoy very difference from standard Baker-Akhiezer functions.
   Hence, one may not expect the usual theta function representations
   of $\psi_j$, $j=1,2,$ in terms of ratios of theta functions
   times a exponential term including $(x-x_0)$ multiplying a
   meromorphic differential with a pole at the essential singularity
   of $\psi_j$. However, using the properties of symmetric
   function and $(F.89)~\cite{15}$, we obtain
   \begin{eqnarray*}
   F_{n+1}(z)&=&z^{n+1}+\sum_{k=0}^{n} \Psi_{n+1-k}(\bar{\mu}) z^k
      \nonumber \\
    &=&
    z^{n+1}+\sum_{k=1}^n \Big( \Psi_{n+1-k}(\underline{\lambda})
      \nonumber
    \end{eqnarray*}
    \begin{eqnarray}\label{4.45}
    &&
    -\sum_{j=1}^nc_j(k) \partial_{\omega_j}
    \mathrm{ln}
    \left(
    \frac{\theta(\underline{z}(P_{\infty_+},\underline{\hat{\mu}})+\underline{\omega})}
    {\theta(\underline{z}(P_{\infty_-},\underline{\hat{\mu}})+\underline{\omega})}
    \right)
    \Big|_{\underline{\omega}=0} \Big) z^{k}
       \nonumber \\
    &=&
      z\prod_{j=1}^n(z-\lambda_j)
       \nonumber \\
     &&
       -\sum_{j=1}^n \sum_{k=1}^n
       c_j(k) \partial_{\omega_j}
    \mathrm{ln}
    \left(
    \frac{\theta(\underline{z}(P_{\infty_+},\underline{\hat{\mu}})+\underline{\omega})}
    {\theta(\underline{z}(P_{\infty_-},\underline{\hat{\mu}})+\underline{\omega})}
    \right)
    \Big|_{\underline{\omega}=0} z^{k},
   \end{eqnarray}
and by inserting $(\ref{4.45})$ into $(\ref{3.26})$, we obtain the
theta function representation of $\psi_1$. Then, the corresponding
theta functions representation of $\psi_2$ follows by $(\ref{3.8})$
and $(\ref{4.34})$.
  \end{rem4.9}

At the end of this section, we turn to the initial value problem in
the stationary case. We show that the solvability of the Dubrovin
equations (\ref{3.38}) on $\Omega_\mu \subseteq \mathbb{R}$ in fact
implies the stationary HS equation (\ref{2.34}) on $\Omega_\mu$,
which amounts to solving the algebro-geometric initial value problem
in the stationary case.

\newtheorem{the4.10}[lem4.1]{Theorem}
 \begin{the4.10}
    Assume that $(\ref{2.1})$  holds and
     $\{\hat{\mu}_j\}_{j=0,\ldots,n}$ satisfies the stationary
    Dubrovin equations $(\ref{3.38})$ on $\Omega_\mu$ and
    remain distinct and nonzero for $ x \in
     \Omega_\mu,$ where $\Omega_\mu \subseteq \mathbb{R}$ is an open
     interval. Then, $u$ defined by
      \begin{equation}\label{4.46}
        u=-\frac{1}{2} \sum_{m=0}^{2n+1} E_m + \frac{1}{2} \sum_{j=0}^n \mu_j,
      \end{equation}
     satisfies the $n$th stationary HS equation
   $(\ref{2.34}),$ that is
     \begin{equation}\label{4.47}
        \mathrm{s}\textrm{-}\mathrm{HS}_n(u)=0,
        \quad \textrm{on $\Omega_\mu.$}
     \end{equation}
  \end{the4.10}
\textbf{Proof.}~~Given the solutions
   $\hat{\mu}_j=(\mu_j,y(\hat{\mu}_j))\in
   C^\infty(\Omega_\mu,\mathcal {K}_n),j=0,\cdots,n$ of (\ref{3.38}),
   let us   introduce
   \begin{equation}\label{4.48}
     F_{n+1}(z)= \prod_{j=0}^n(z-\mu_j)
     \quad \textrm{on $\mathbb{C}\times\Omega_\mu$},
    \end{equation}
 with $u$ defined by (\ref{4.46}) up to multiplicative
 constant. Given $F_{n+1}$ and $u$, let us denote the polynomial $G_n$ by
    \begin{equation}\label{4.49}
        G_n(z)=\frac{1}{2}F_{n+1,x}(z),
        \quad
         \textrm{on $\mathbb{C}\times\Omega_\mu$},
    \end{equation}
 and from ({\ref{4.48}), one can see that the degree of $G_n$ is $n$
 with respect to $z$. Taking account into (\ref{4.48}),
 the Dubrovin equations (\ref{3.38}) imply
     \begin{equation}\label{4.50}
       y(\hat{\mu}_j)= \frac{1}{2} \mu_j \mu_{j,x}
         \prod_{ \scriptstyle k=0 \atop \scriptstyle k \neq j}^n
         (\mu_j-\mu_k)
       = -\frac{1}{2} \mu_j F_{n+1,x}(\mu_j)
       =-\mu_j G_n(\mu_j).
      \end{equation}
  Hence
     \begin{equation}\label{4.51}
        R_{2n+2}(\mu_j)^2-\mu_j^2 G_n(\mu_j)^2=
        y(\hat{\mu}_j)^2-\mu_j^2 G_n(\mu_j)^2=0,
        \quad j=0,\ldots,n.
     \end{equation}
 Next, let us define a polynomial $H_n$ on $\mathbb{C}\times\Omega_\mu$
 such that
    \begin{equation}\label{4.52}
      R_{2n+2}(z)-z^2G_n(z)^2=zF_{n+1}(z)H_n(z)
    \end{equation}
 holds. Such a polynomial $H_n$ exists since the left-hand side of (\ref{4.52})
 vanishes at $z=\mu_j,~j=0,\cdots,n$ by (\ref{4.51}). We need to
 determine the degree of $H_n$. By (\ref{4.49}), we compute
   \begin{equation}\label{4.53}
     R_{2n+2}(z)-z^2G_n(z)^2 \underset{|z| \rightarrow \infty}{=}
    h_0 z^{2n+2}+O(z^{2n+1}),
   \end{equation}
 with $O(z^{2n+1})$ depending on $x$ by
 inspection. Therefore, combining (\ref{4.48}), (\ref{4.49}),
 (\ref{4.52}) and (\ref{4.53}), we conclude that $H_n$ has degree $n$ with respect to
 $z$, with the coefficient $h_0$ of powers $z^{n}$. Hence, we
 may write $H_n$ as
      \begin{equation}\label{4.54}
        H_n(z)=h_0 \prod_{l=1}^n
        (z-\nu_l),
         \quad
         \textrm{on $\mathbb{C}\times\Omega_\mu$}.
     \end{equation}
 Next, let us consider the polynomial $P_{n}$ by
    \begin{equation}\label{4.55}
        P_{n}(z)=H_n(z)+u_{xx}F_{n+1}(z)
        +zG_{n,x}(z).
    \end{equation}
 Using (\ref{4.48}), (\ref{4.49}) and (\ref{4.54}) we obtain that
 $P_{n}$ is a polynomial of degree at most $n$. Differentiating on both sides of
 (\ref{4.52}) with respect to $x$ yields
     \begin{equation}\label{4.56}
        2z^2G_n(z)G_{n,x}(z)+zF_{n+1,x}(z)H_n(z)+zF_{n+1}(z)H_{n,x}(z)=0
        \quad \textrm{on $\mathbb{C}\times\Omega_\mu$}.
    \end{equation}
 Multiplying (\ref{4.55}) by $G_n$ and using (\ref{4.56}), we have
    \begin{eqnarray}\label{4.57}
     G_n(z)P_{n}(z)&=& F_{n+1}(z)(u_{xx}G_n(z)-\frac{1}{2}H_{n,x}(z))
       \nonumber \\
     &&+~
     (G_n(z)-\frac{1}{2}F_{n+1,x}(z)) H_n(z),
  \end{eqnarray}
 and hence
    \begin{equation}\label{4.58}
        G_n(\mu_j)P_{n}(\mu_j)=0,
        \qquad j=1,\ldots,n,
    \end{equation}
 on $\Omega_\mu$ by using (\ref{4.49}).\\
 Next, let
   $x\in\widetilde{\Omega}_\mu \subseteq \Omega_\mu$,
 where $\widetilde{\Omega}_\mu$ is given by
      \begin{eqnarray}\label{4.59}
        \widetilde{\Omega}_\mu&=&\{x\in\Omega_\mu \mid
        G(\mu_j(x),x)=-\frac{y(\hat{\mu}_j(x))}{\mu_j(x)}\neq 0,\,j=0,\cdots,n\}
        \nonumber \\
        &=&\{x\in\Omega_\mu \mid \mu_j(x)
        \notin \{E_m \}_{m=0,\cdots,2n+1},j=0,\cdots,n\},
      \end{eqnarray}
 Thus, we have
      \begin{equation}\label{4.60}
        P_{n}(\mu_j(x),x)=0,\quad
        j=0,\cdots,n,\, ~ x\in\widetilde{\Omega}_\mu.
      \end{equation}
 Since $P_{n}$ is a polynomial of degree at most $n$, (\ref{4.60})
 implies
     \begin{equation}\label{4.61}
        P_{n}=0
        \quad \textrm{on $\mathbb{C}\times\widetilde{\Omega}_\mu$},
     \end{equation}
 So, (2.17) holds, that is,
       \begin{equation}\label{4.62}
        zG_{n,x}(z)=-H_n(z)-u_{xx}F_{n+1}(z)
         \quad \textrm{on $\mathbb{C}\times\widetilde{\Omega}_\mu$}.
       \end{equation}
 Inserting (\ref{4.62}) and (\ref{4.49}) into (\ref{4.56}) yields
         \begin{equation}\label{4.63}
             zF_{n+1}(z) (-2u_{xx}G_n(z)+H_{n,x}(z))=0,
         \end{equation}
 namely
       \begin{equation}\label{4.64}
        H_{n,x}(z)=2u_{xx}G_n(z),
         \quad \textrm{on $\mathbb{C}\times\widetilde{\Omega}_\mu$}.
       \end{equation}
 Thus, we obtain the fundamental equations (2.15)-(2.17), and (\ref{2.19})
 on $\mathbb{C}\times\widetilde{\Omega}_\mu$.

 In order to extend these results to all $x\in\Omega_\mu$, let us
 consider the case where $\hat{\mu}_j$ admits one of the
 branch points $(E_{m_0},0)$. Hence, we suppose
   \begin{equation}\label{4.65}
      \mu_{j_1}(x)\rightarrow E_{m_0}
      \quad \textrm{as $x\rightarrow x_0\in\Omega_\mu$},
   \end{equation}
 for some $j_1\in\{0,\cdots,n\},\,m_0\in\{1,\cdots,2n+1\}$.
 Introducing
 \begin{equation}\label{4.66}
   \begin{split}
 & \zeta_{j_1}(x)=\sigma(\mu_{j_1}(x)-E_{m_0})^{1/2},
   \quad \sigma=\pm1,\quad \\
  & \mu_{j_1}(x)=E_{m_0}+\zeta_{j_1}(x)^2
    \end {split}
 \end{equation}
 for some $x$ in an open interval centered near $x_0$,
 then the Dubrovin
 equation (\ref{3.38}) for $\mu_{j_1}$ becomes
   \begin{eqnarray}\label{4.67}
 \zeta_{j_1,x}(x)&=&
 c(\sigma) \frac{a }{E_{m_0}}
 \Big(\prod_{\scriptstyle m=0 \atop \scriptstyle m\neq
  m_0}^{2n+1}(E_{m_0}-E_m)\Big)^{1/2} \nonumber \\
   &&\times \prod_{\scriptstyle k=0 \atop \scriptstyle k\neq
   j_1}^{n}(E_{m_0}-\mu_k(x))^{-1}(1+O(
   \zeta_{j_1}(x)^2))
 \end{eqnarray}
  for some $| c(\sigma)|=1$. Hence (\ref{4.61})-(\ref{4.64}) extend to
  $\Omega_\mu$ by continuity. Consequently, we obtain relations
  (2.15)-(2.17) on $\mathbb{C}\times \Omega_\mu$, and can proceed
  as in Section 2 to see that $u$ satisfies the stationary HS hierarchy
  (\ref{4.47}). \quad $\square$

\newtheorem{rem4.11}[lem4.1]{Remark}
  \begin{rem4.11}
  The result in Theorem $4.10$ is derived in terms of
  $u$ and $\{\mu_j\}_{j=0,\cdots,n}$, but
  one can prove the analogous result
  in terms of $u$ and $\{\nu_l\}_{l=1,\cdots,n}$.
  \end{rem4.11}
\newtheorem{rem4.12}[lem4.1]{Remark}
   \begin{rem4.12}
    Theorem $4.10$ reveals that given $\mathcal{K}_n$ and the initial
    condition $(\hat{\mu}_0(x_0),\hat{\mu}_1(x_0),\ldots,\hat{\mu}_n(x_0))
    $, or equivalently, the
    auxiliary divisor $\mathcal{D}_{\hat{\mu}_0(x_0) \underline{\hat{\mu}}(x_0)}
    $ at $x=x_0$, $u$ is uniquely
    determined in an open neighborhood $\Omega$ of $x_0$
    by $(\ref{4.46})$ and satisfies the $n$th stationary HS equation
    $(\ref{2.34})$. Conversely, given $\mathcal{K}_n$
    and $u$ in an open neighborhood $\Omega$ of $x_0$, we can
    construct the corresponding polynomial $F_{n+1}(z,x)$, $G_n(z,x)$
    and $H_n(z,x)$ for $x\in\Omega$, and then obtain the auxiliary
    divisor $\mathcal{D}_{\hat{\mu}_0(x) \underline{\hat{\mu}}(x)}$ for
    $x\in\Omega$ from the zeros of $F_{n+1}(z,x)$ and $(\ref{3.10})$.
    In that sense, once the curve $\mathcal{K}_n$ is fixed, elements of the
    isospectral class of the HS potentials $u$ can be characterized by
    nonspecial auxiliary divisor
    $\mathcal{D}_{\hat{\mu}_0(x) \underline{\hat{\mu}}(x)}$.
   \end{rem4.12}

\section{The time-dependent HS formalism}
 In this section, let us go back to the recursive approach detailed in
 Section 2 and
 extend the the algebro-geometric analysis of Section 3 to the
 time-dependent HS hierarchy.

 Throughout this section we assume (\ref{2.2}) to hold.

 The time-dependent algebro-geometric initial value problem of the
 HS hierarchy is to solve the time-dependent $r$th HS flow with
 a stationary solution of the $n$th equation as initial data in the
 hierarchy. More precisely, given $n\in\mathbb{N}_0$, based on the
 solution $u^{(0)}$ of the $n$th stationary HS equation
 $\textrm{s-HS}_n(u^{(0)})=0$ associated with $\mathcal{K}_n$ and a
 set of integration constants $\{c_l\}_{l=1,\ldots,n} \subset
 \mathbb{C}$, we want to build up  a solution $u$ of the $r$th HS flow
 $\mathrm{HS}_r(u)=0$ such that $u(t_{0,r})=u^{(0)}$ for some
 $t_{0,r} \in \mathbb{R},~r\in\mathbb{N}_0$.

 We employ the notations
 $\widetilde{V}_r,$ $\widetilde{F}_{r+1},$ $\widetilde{G}_r,$
 $\widetilde{H}_r,$ $\tilde{f}_{s}$, $\tilde{g}_{s},$ $\tilde{h}_{s}$
 to stand for the time-dependent quantities, which are obtained in
 $V_n,$ $F_{n+1},$ $G_n,$ $H_n,$ $f_{l},$ $g_{l},$ $h_{l}$ by
 replacing $\{c_l\}_{l=1,\ldots,n}$ with
 $\{\tilde{c}_s\}_{s=1,\ldots,r}$, where the integration constants
 $\{c_l\}_{l=1,\ldots,n}\subset \mathbb{C}$ in the stationary HS hierarchy and
 $\{\tilde{c}_s\}_{s=1,\ldots,r}\subset \mathbb{C}$ in the time-dependent HS
 hierarchy are independent of each other. In addition, we mark
 the individual $r$th HS flow by a separate time variable $t_r \in
 \mathbb{R}$.

 Let us now  provide the time-dependent
 algebro-geometric initial value problem as follows
   \begin{eqnarray}\label{5.1}
    &&
     \begin{split}
       & \mathrm{HS}_r(u)=-u_{xxt_r}+u_{xxx}\tilde{f}_{r+1}(u)
          +2u_{xx}\tilde{f}_{r+1,x}(u)=0, \\
       & u|_{t_r=t_{0,r}}=u^{(0)},
      \end{split}
          \\
   &&
      \textrm{s-HS}_n(u^{(0)})=u_{xxx}f_{n+1}(u^{(0)})
      +2u_{xx}f_{n+1,x}(u^{(0)})=0,
   \end{eqnarray}
 where $t_{0,r}\in\mathbb{R},$ $n,r\in\mathbb{N}_0$, $u=u(x,t_r)$
 satisfies the condition (\ref{2.2}), and the curve $\mathcal{K}_n$ is
 associated with the initial data $u^{(0)}$ in (5.2). Noticing that
 the HS flows are isospectral, we are going a further step and
 assume that (5.2) holds not only at $t_r=t_{0,r}$, but also at all $t_r \in
 \mathbb{R}$.

 Let us now start from the zero-curvature equations (\ref{2.41})
    \begin{equation}\label{5.3}
        U_{t_r}-\widetilde{V}_{r,x}+[U,\widetilde{V}_r]=0,
    \end{equation}
    \begin{equation}\label{5.4}
        -V_{n,x}+[U,V_n]=0,
    \end{equation}
 where
 \begin{equation}\label{5.5}
    \begin{split}
    & U(z)=
     \left(
       \begin{array}{cc}
         0 & 1 \\
         -z^{-1}u_{xx} & 0 \\
       \end{array}
     \right)
     \\
    & V_n(z)=
    \left(
      \begin{array}{cc}
      -G_n(z) & F_{n+1}(z) \\
      z^{-1}H_n(z) & G_n(z) \\
      \end{array}
    \right)
      \\
    & \widetilde{V}_r(z)=
       \left(
         \begin{array}{cc}
           -\widetilde{G}_r(z) & \widetilde{F}_{r+1}(z) \\
           z^{-1}\widetilde{H}_r(z) & \widetilde{G}_r(z) \\
         \end{array}
       \right)
    \end{split}
  \end{equation}
and
  \begin{eqnarray}\label{5.6}
   &&
    F_{n+1}(z)=\sum_{l=0}^{n+1} f_{l} z^{n+1-l}
    =\prod_{j=0}^n (z-\mu_j),
    \\
   &&
    G_n(z)=\sum_{l=0}^n g_{l} z^{n-l},
     \\
   &&
    H_n(z)=\sum_{l=0}^n h_{l} z^{n-l}
    =h_0 \prod_{l=1}^n (z-\nu_l),
    \end{eqnarray}
 \begin{eqnarray}
 &&
    \widetilde{F}_{r+1}(z)=\sum_{s=0}^{r+1} \tilde{f}_{s} z^{r+1-s},
    \\
   &&
    \widetilde{G}_r(z)=\sum_{s=0}^r \tilde{g}_{s} z^{r-s},
    \\
    &&
    \widetilde{H}_r(z)=\sum_{s=0}^r \tilde{h}_{s} z^{r-s},
   \end{eqnarray}
 for fixed $n,r\in \mathbb{N}_0$. Here $\{f_{l}\}_{l=0,\ldots,n+1},$
 $\{g_{l}\}_{l=0,\ldots,n}$, $\{h_{l}\}_{l=0,\ldots,n}$,
 and $\{\tilde{f}_{s}\}_{s=0,\ldots,r+1},$
 $\{\tilde{g}_{s}\}_{s=0,\ldots,r}$, $\{\tilde{h}_{s}\}_{s=0,\ldots,r}$,
 satisfy the relations in (\ref{2.3}).

 Moreover, it is more convenient for us to rewrite the
 zero-curvature equations (\ref{5.3}) and (\ref{5.4}) as the
 following forms,
    \begin{equation}\label{5.12}
        -u_{xxt_r}-\widetilde{H}_{r,x}
        +2u_{xx}\widetilde{G}_r=0,
    \end{equation}
     \begin{equation}\label{5.13}
        \widetilde{F}_{r+1,x}
        =2\widetilde{G}_r,
    \end{equation}
    \begin{equation}\label{5.14}
        z\widetilde{G}_{r,x}=-\widetilde{H}_r
         -u_{xx}\widetilde{F}_{r+1}
    \end{equation}
and
   \begin{equation}\label{5.15}
     F_{n+1,x}=2G_n,
   \end{equation}
   \begin{equation}\label{5.16}
     H_{n,x}=2u_{xx}G_n,
   \end{equation}
   \begin{equation}\label{5.17}
        zG_{n,x}=-H_n -u_{xx}F_{n+1}.
    \end{equation}
From (\ref{5.15})-(\ref{5.17}), we may compute
  \begin{equation}\label{5.18}
    \frac{d}{dx} \mathrm{det}(V_n(z))=-\frac{1}{z^2}\frac{d}{dx}
    \Big( z^2G_n(z)^2+zF_{n+1}(z)H_n(z) \Big)=0,
  \end{equation}
and meanwhile Lemma 5.2 gives
  \begin{equation}\label{5.19}
    \frac{d}{dt_r} \mathrm{det}(V_n(z))=-\frac{1}{z^2}\frac{d}{dt_r}
    \Big( z^2G_n(z)^2+zF_{n+1}(z)H_n(z) \Big)=0,
  \end{equation}
Hence, $z^2G_n(z)^2+zF_{n+1}(z)H_n(z)$ is independent of variables
both $x$ and $t_r$, which implies
   \begin{equation}\label{5.20}
   z^2G_n(z)^2+zF_{n+1}(z)H_n(z)=R_{2n+2}(z).
   \end{equation}
This reveals that the fundamental identity (\ref{2.19}) still holds
in the time-dependent context. Consequently the hyperelliptic curve
$\mathcal{K}_n$ is still available by (\ref{2.27}).

Next, let us introduce the time-dependent Baker-Akhiezer function
$\psi(P,x,x_0,$ $t_r,t_{0,r})$  on $\mathcal{K}_{n}\setminus
\{P_{\infty_\pm},P_0\}$ by
  \begin{equation}\label{5.21}
         \begin{split}
          & \psi(P,x,x_0,t_r,t_{0,r})=\left(
                            \begin{array}{c}
                              \psi_1(P,x,x_0,t_r,t_{0,r}) \\
                              \psi_2(P,x,x_0,t_r,t_{0,r}) \\
                            \end{array}
                          \right), \\
          & \psi_x(P,x,x_0,t_r,t_{0,r})=U(u(x,t_r),z(P))\psi(P,x,x_0,t_r,t_{0,r}),\\
          &  \psi_{t_r}(P,x,x_0,t_r,t_{0,r})=\widetilde{V}_r(u(x,t_r),z(P))
               \psi(P,x,x_0,t_r,t_{0,r}), \\
           & zV_n(u(x,t_r),z(P))\psi(P,x,x_0,t_r,t_{0,r})
           =y(P)\psi(P,x,x_0,t_r,t_{0,r}),\\
          & \psi_1(P,x_0,x_0,t_{0,r},t_{0,r})=1; \\
          &
          P=(z,y)\in \mathcal{K}_{n}
           \setminus \{P_{\infty_\pm},P_0\},~(x,t_r)\in
           \mathbb{R}^2,
         \end{split}
   \end{equation}
where
   \begin{eqnarray}\label{5.22}
   \psi_1(P,x,x_0,t_r,t_{0,r})&=&\mathrm{exp}\Big(
     \int_{t_{0,r}}^{t_r} ds
     (z^{-1}\widetilde{F}_{r+1}(z,x_0,s)\phi(P,x_0,s)
     \nonumber \\
     &&
     -\widetilde{G}_r(z,x_0,s))
     + z^{-1} \int_{x_0}^x dx^\prime
     \phi(P,x^\prime,t_r)
      \Big), \nonumber \\
     && ~~~~~~~~~~~
     P=(z,y)\in \mathcal{K}_{n}
           \setminus \{P_{\infty_\pm},P_0\}.
   \end{eqnarray}
Closely related to $\psi(P,x,x_0,t_r,t_{0,r})$ is the following
meromorphic function $\phi(P,x,t_r)$ on $\mathcal{K}_{n}$ defined by
        \begin{equation}\label{5.23}
         \phi(P,x,t_r)=z
         \frac{ \psi_{1,x}(P,x,x_0,t_r,t_{0,r})}
          {\psi_1(P,x,x_0,t_r,t_{0,r})},
                      \quad P \in
          \mathcal{K}_{n}\setminus \{P_{\infty_\pm},P_0\},
          ~ (x,t_r)\in \mathbb{R}^2.
        \end{equation}
which implies by (\ref{5.21}) that
   \begin{eqnarray}\label{5.24}
        \phi(P,x,t_r)&=& \frac{y+zG_n(z,x,t_r)}{F_{n+1}(z,x,t_r)}
           \nonumber \\
        &=&
         \frac{zH_n(z,x,t_r)}{y-zG_n(z,x,t_r)},
    \end{eqnarray}
 and
      \begin{equation}\label{5.25}
        \psi_2(P,x,x_0,t_r,t_{0,r})=
        \psi_1(P,x,x_0,t_r,t_{0,r})\phi(P,x,t_r)/z.
      \end{equation}
In analogy to equations (\ref{3.10}) and (\ref{3.11}), we define
     \begin{eqnarray}\label{5.26}
       \hat{\mu}_j(x,t_r)=(\mu_j(x,t_r),
       -\mu_j(x,t_r)G_n(\mu_j(x,t_r),x,t_r))
       \in \mathcal{K}_n,
        \nonumber \\
        j=0,\ldots,n, ~(x,t_r)\in \mathbb{R}^2,
     \end{eqnarray}
     \begin{eqnarray}\label{5.27}
       \hat{\nu}_l(x,t_r)=(\nu_l(x,t_r),
       \nu_l(x,t_r)G_n(\nu_l(x,t_r),x,t_r))
       \in \mathcal{K}_n,
        \nonumber \\
        l=1,\ldots,n, ~(x,t_r)\in \mathbb{R}^2.
     \end{eqnarray}
The regular properties of $F_{n+1}$, $H_n$, $\mu_j$ and $\nu_l$ are
analogous to those in Section 3 due to assumptions
(\ref{2.2}).\\
From (\ref{5.24}), the the divisor $(\phi(P,x,t_r))$ of
$\phi(P,x,t_r)$ reads
 \begin{equation}\label{5.28}
           (\phi(P,x,t_r))=
           \mathcal{D}_{P_0\underline{\hat{\nu}}(x,t_r)}(P)
           -\mathcal{D}_{\hat{\mu}_0(x,t_r)\underline{\hat{\mu}}(x,t_r)}(P)
  \end{equation}
where
  \begin{equation}\label{5.29}
    \underline{\hat{\mu}}=\{\hat{\mu}_1,\ldots,\hat{\mu}_{n}\},
    \quad
    \underline{\hat{\nu}}=\{\hat{\nu}_1,\ldots,\hat{\nu}_{n}\}
    \in \mathrm{Sym}^n (\mathcal{K}_n).
  \end{equation}
That means $P_0,\hat{\nu}_1(x,t_r),\ldots,\hat{\nu}_{n}(x,t_r)$ are
the $n+1$ zeros of $\phi(P,x,t_r)$ and
$\hat{\mu}_0(x,t_r),\hat{\mu}_1(x,t_r), \ldots,\hat{\mu}_{n}(x,t_r)$
its $n+1$ poles.\\

Further properties of $\phi(P,x,t_r)$ are summarized as follows.

\newtheorem{lem5.1}{Lemma}[section]
 \begin{lem5.1}
  Assume that $(\ref{2.2})$, $(\ref{5.3})$ and
    $(\ref{5.4})$ hold. Let
    $P=(z,y) \in \mathcal{K}_{n}\setminus \{P_{\infty_\pm},P_0\},
     ~ (x,t_r)\in \mathbb{R}^2.$ Then
  \begin{equation}\label{5.30}
    \phi_x(P)+ z^{-1}\phi(P)^2 =-u_{xx},
  \end{equation}
  \begin{eqnarray}\label{5.31}
        \phi_{t_r}(P)&=&(-z\widetilde{G}_r(z)+\widetilde{F}_{r+1}(z)\phi(P))_x
            \\
        &=&\widetilde{H}_r(z)+u_{xx}\widetilde{F}_{r+1}(z)
        +(\widetilde{F}_{r+1}(z)\phi(P))_x, \nonumber
  \end{eqnarray}
  \begin{equation}\label{5.32}
        \phi_{t_r}(P)=\widetilde{H}_r(z)+2\widetilde{G}_r(z)\phi(P)
         -z^{-1}\widetilde{F}_{r+1}(z)\phi(P)^2,
  \end{equation}
  \begin{equation}\label{5.33}
    \phi(P)\phi(P^\ast)=-\frac{zH_n(z)}{F_{n+1}(z)},
  \end{equation}
  \begin{equation}\label{5.34}
    \phi(P)+\phi(P^\ast)=2\frac{zG_n(z)}{F_{n+1}(z)},
  \end{equation}
  \begin{equation}\label{5.35}
    \phi(P)-\phi(P^\ast)=\frac{2y}{F_{n+1}(z)}.
  \end{equation}
 \end{lem5.1}
\textbf{Proof.}~~We just need to prove (\ref{5.31}) and
(\ref{5.32}). Equations (\ref{5.30}) and (\ref{5.33})-(\ref{5.35})
can be proved as in Lemma 3.1. By using (\ref{5.21}) and
(\ref{5.23}), we obtain
 \begin{eqnarray}\label{5.36}
        \phi_{t_r}&=&
        z(\mathrm{ln}\psi_1)_{xt_r}=z(\mathrm{ln}\psi_1)_{t_rx}
        =z\Big(\frac{\psi_{1,t_r}}{\psi_1}\Big)_x
          \nonumber \\
        &=& z
 \Big(\frac{-\widetilde{G}_r\psi_1+\widetilde{F}_{r+1}\psi_2}{\psi_1}\Big)_x
          \nonumber \\
      &=&
      (-z\widetilde{G}_r+\widetilde{F}_{r+1}\phi)_x,
 \end{eqnarray}
which is the fist line of (\ref{5.31}). Inserting (\ref{5.14}) into
(\ref{5.36}) yields the second line of (\ref{5.31}). Then by the
definition of $\phi$ (\ref{5.23}), one may have
 \begin{eqnarray}\label{5.37}
     \phi_{t_r} &=& z \Big(\frac{\psi_2}{\psi_1}\Big)_{t_r}
     =
      z \Big(\frac{\psi_{2,t_r}}{\psi_1}-
      \frac{\psi_2\psi_{1,t_r}}{\psi_1^2} \Big)
        \nonumber \\
     &=&
     z \Big(\frac{z^{-1}\widetilde{H}_r\psi_1+\widetilde{G}_r\psi_2}{\psi_1}
     -z^{-1}\phi\frac{-\widetilde{G}_r\psi_1+\widetilde{F}_{r+1}\psi_2}{\psi_1}\Big)
       \nonumber \\
     &=&
     \widetilde{H}_r+2\widetilde{G}_r\phi-z^{-1}\widetilde{F}_{r+1}\phi^2,
 \end{eqnarray}
which is (\ref{5.32}). Alternatively, one can insert (5.12)-(5.14)
into (\ref{5.31}) to obtain (\ref{5.32}). \quad $\square$ \\

Next we study the time evolution of $F_{n+1}$, $G_n$ and $H_n$ by
using zero-curvature equations (5.12)-(5.14) and (5.15)-(5.17).

\newtheorem{lem5.2}[lem5.1]{Lemma}
 \begin{lem5.2}
  Assume that $(\ref{2.2})$ ,$(\ref{5.3})$ and
  $(\ref{5.4})$ hold. Then
   \begin{equation}\label{5.38}
    F_{n+1,t_r}=2(G_n\widetilde{F}_{r+1}-\widetilde{G}_rF_{n+1}),
   \end{equation}
   \begin{equation}\label{5.39}
    zG_{n,t_r}=\widetilde{H}_rF_{n+1}-H_n\widetilde{F}_{r+1},
   \end{equation}
    \begin{equation}\label{5.40}
        H_{n,t_r}=2(H_n\widetilde{G}_r-G_n\widetilde{H}_r).
    \end{equation}
 Equations $(\ref{5.38})-(\ref{5.40})$ imply
    \begin{equation}\label{5.41}
        -V_{n,t_r}+[\widetilde{V}_r,V_n]=0.
    \end{equation}
 \end{lem5.2}
\textbf{Proof.}~~Differentiating both sides of (\ref{5.35}) with
respect to $t_r$ leads to
  \begin{equation}\label{5.42}
    (\phi(P)-\phi(P^\ast))_{t_r}=-2yF_{n+1,t_r}F_{n+1}^{-2}.
  \end{equation}
On the other hand, by (\ref{5.32}), (\ref{5.34}) and (\ref{5.35}),
the left-hand side of (\ref{5.42}) equals to
   \begin{eqnarray}\label{5.43}
    \phi(P)_{t_r}-\phi(P^\ast)_{t_r}&=&
    2\widetilde{G}_r(\phi(P)-\phi(P^\ast))-z^{-1}\widetilde{F}_{r+1}
    (\phi(P)^2-\phi(P^\ast)^2)
     \nonumber \\
    &=&
    4y(\widetilde{G}_rF_{n+1}-\widetilde{F}_{r+1}G_n)F_{n+1}^{-2}.
   \end{eqnarray}
Combining (\ref{5.42}) with (\ref{5.43}) yields (\ref{5.38}).
Similarly, Differentiating both sides of (\ref{5.34}) with respect
to $t_r$ gives
  \begin{equation}\label{5.44}
    (\phi(P)+\phi(P^\ast))_{t_r}=2z(G_{n,t_r}F_{n+1}-G_nF_{n+1,t_r})F_{n+1}^{-2},
  \end{equation}
Meanwhile, by (\ref{5.32}), (\ref{5.33}) and (\ref{5.34}), the
left-hand side of (\ref{5.44})  equals to
  \begin{eqnarray}\label{5.45}
  \phi(P)_{t_r}+\phi(P^\ast)_{t_r}&=&
   2\widetilde{G}_r(\phi(P)+\phi(P^\ast))
   -z^{-1}\widetilde{F}_{r+1}(\phi(P)^2+\phi(P^\ast)^2)+2\widetilde{H}_r
    \nonumber \\
   &=&
   -2zG_nF_{n+1}^{-2}F_{n+1,t_r}
   +2F_{n+1}^{-1}(\widetilde{H}_rF_{n+1}-\widetilde{F}_{r+1}H_n).
  \end{eqnarray}
 Thus, (\ref{5.39})
clearly  follows by (\ref{5.44}) and (\ref{5.45}). Hence, insertion
of (\ref{5.38}) and (\ref{5.39}) into the differentiation of
$z^2G_n^2+zF_{n+1}H_n=R_{2n+2}(z)$ can derive (\ref{5.40}). Finally,
a direct calculation shows that (\ref{5.38})-(\ref{5.40})
are equivalent to (\ref{5.41}). \quad $\square$ \\

Further properties of $\psi$ are summarized as follows.

\newtheorem{lem5.3}[lem5.1]{Lemma}
 \begin{lem5.3}
  Assume that $(\ref{2.2})$, $(\ref{5.3})$ and
  $(\ref{5.4})$ hold. Let
  $P=(z,y) \in \mathcal{K}_{n}\setminus \{P_{\infty_\pm},P_0\},
   ~ (x,x_0,t_r.t_{0,r})\in \mathbb{R}^4.$
  Then, we have
  \begin{eqnarray}\label{5.46}
    \psi_1(P,x,x_0,t_r,t_{0,r}) &=&
    \Big(\frac{F_{n+1}(z,x,t_r)}{F_{n+1}(z,x_0,t_{0,r})}\Big)^{1/2}
     \nonumber \\
     &\times&
    \mathrm{exp} \Big(
    \frac{y}{z}\int_{t_{0,r}}^{t_r} ds
    \widetilde{F}_{r+1}(z,x_0,s)F_{n+1}(z,x_0,s)^{-1}
    \nonumber \\
    &&
    +\frac{y}{z} \int_{x_0}^x dx^\prime
    F_{n+1}(z,x^\prime,t_r)^{-1}
    \Big),
  \end{eqnarray}
  \begin{equation}\label{5.47}
    \psi_1(P,x,x_0,t_r,t_{0,r}) \psi_1(P^\ast,x,x_0,t_r,t_{0,r})
    =\frac{F_{n+1}(z,x,t_r)}{F_{n+1}(z,x_0,t_{0,r})},
  \end{equation}
  \begin{equation}\label{5.48}
    \psi_2(P,x,x_0,t_r,t_{0,r}) \psi_2(P^\ast,x,x_0,t_r,t_{0,r})
    =-\frac{H_n(z,x,t_r)}{z F_{n+1}(z,x_0,t_{0,r})},
  \end{equation}
   \begin{eqnarray}\label{5.49}
  &&
    \psi_1(P,x,x_0,t_r,t_{0,r}) \psi_2(P^\ast,x,x_0,t_r,t_{0,r})
      \\
  && ~~~~
    + \psi_1(P^\ast,x,x_0,t_r,t_{0,r}) \psi_2(P,x,x_0,t_r,t_{0,r})
    =2\frac{G_n(z,x,t_r)}{F_{n+1}(z,x_0,t_{0,r})},
    \nonumber
  \end{eqnarray}
   \begin{eqnarray}\label{5.50}
  &&
    \psi_1(P,x,x_0,t_r,t_{0,r}) \psi_2(P^\ast,x,x_0,t_r,t_{0,r})
      \\
  && ~~~~
    - \psi_1(P^\ast,x,x_0,t_r,t_{0,r}) \psi_2(P,x,x_0,t_r,t_{0,r})
    =-\frac{2y}{z F_{n+1}(z,x_0,t_{0,r})}.
    \nonumber
  \end{eqnarray}
 \end{lem5.3}
\textbf{Proof.}~~In order to prove (\ref{5.46}), let us first
consider the part of time variable in the definition (\ref{5.22}),
that is
   \begin{equation}\label{5.51}
    \mathrm{exp} \Big( \int_{t_{0,r}}^{t_r} ds~
    (z^{-1}\widetilde{F}_{r+1}(z,x_0,s)\phi(P,x_0,s)
    -\widetilde{G}_r(z,x_0,s))
    \Big).
   \end{equation}
The integrand in the above integral equals to
   \begin{eqnarray}\label{5.52}
   &&
     z^{-1}\widetilde{F}_{r+1}(z,x_0,s)\phi(P,x_0,s)
     -\widetilde{G}_r(z,x_0,s)
      \nonumber \\
     &&
     =z^{-1}\widetilde{F}_{r+1}(z,x_0,s)
     \frac{y+zG_n(z,x_0,s)}{F_{n+1}(z,x_0,s)}
     -\widetilde{G}_r(z,x_0,s)
       \nonumber \\
     &&
     =\frac{y}{z}\widetilde{F}_{r+1}(z,x_0,s)F_{n+1}(z,x_0,s)^{-1}
      +(\widetilde{F}_{r+1}(z,x_0,s)G_n(z,x_0,s)
      \nonumber \\
     && ~~~~
     -\widetilde{G}_r(z,x_0,s)F_{n+1}(z,x_0,s))F_{n+1}(z,x_0,s)^{-1}
       \nonumber \\
      &&
      =\frac{y}{z}\widetilde{F}_{r+1}(z,x_0,s)F_{n+1}(z,x_0,s)^{-1}
      +\frac{1}{2}\frac{F_{n+1,s}(z,x_0,s)}{F_{n+1}(z,x_0,s)},
   \end{eqnarray}
where we used (\ref{5.24}) and (\ref{5.38}). By (\ref{5.52}),
(\ref{5.51}) reads
   \begin{equation}\label{5.53}
   \Big(\frac{F_{n+1}(z,x_0,t_r)}{F_{n+1}(z,x_0,t_{0,r})}\Big)^{1/2}
     \mathrm{exp} \Big( \frac{y}{z} \int_{t_{0,r}}^{t_r} ds
     \widetilde{F}_{r+1}(z,x_0,s)F_{n+1}(z,x_0,s)^{-1}
     \Big).
   \end{equation}
 On the other hand, the
part of space variable in (\ref{5.22}) can be written as
 \begin{equation}\label{5.54}
    \Big(\frac{F_{n+1}(z,x,t_r)}{F_{n+1}(z,x_0,t_{r})}\Big)^{1/2}
    \mathrm{exp}\Big(\frac{y}{z} \int_{x_0}^x dx^\prime
    F_{n+1}(z,x^\prime,t_r)^{-1}  \Big),
 \end{equation}
which can be proved using the similar procedure to Lemma 3.2.
Combining (\ref{5.53}) and (\ref{5.54}) yields (\ref{5.46}).
Evaluating (\ref{5.46}) at the points $P$ and $P^\ast$, and
multiplying the resulting expressions, with noticing
  \begin{equation}\label{5.55}
        y(P)+y(P^\ast)=0,
    \end{equation}
leads to (\ref{5.47}). The remaining statements
(\ref{5.48})-(\ref{5.50}) are direct consequence of (\ref{5.25}),
(\ref{5.33})-(\ref{5.35}) and (\ref{5.47}). \quad $\square$ \\

In analogy to Lemma 3.4, the dynamics of the zeros
$\{\mu_j(x,t_r)\}_{j=0,\ldots,n}$ and
$\{\nu_l(x,t_r)\}_{l=1,\ldots,n}$ of $F_{n+1}(z,x,t_r)$ and
$H_n(z,x,t_r)$ with respect to $x$ and $t_r$ are described in terms
of Dubrovin-type equations (see the following Lemma). We assume that
the affine part of $\mathcal{K}_n$ to be nonsingular, which implies
(\ref{3.37}) holds in present context.

\newtheorem{lem5.4}[lem5.1]{Lemma}
 \begin{lem5.4}
   Assume that $(\ref{2.2})$, $(\ref{5.3})$ and
   $(\ref{5.4})$ hold.

   $(\mathrm{i})$
 Suppose that the zeros $\{\mu_j(x,t_r)\}_{j=0,\ldots,n}$
 of $F_{n+1}(z,x,t_r)$ remain distinct for $(x,t_r) \in
 \Omega_\mu,$ where $\Omega_\mu \subseteq \mathbb{R}^2$ is open
 and connected. Then, $\{\mu_j(x,t_r)\}_{j=0,\ldots,n}$ satisfy
 the system of differential equations,
    \begin{equation}\label{5.56}
        \mu_{j,x}=2\frac{y(\hat{\mu}_j)}{\mu_j}
        \prod_{ \scriptstyle k=0 \atop \scriptstyle k \neq j}^n
        (\mu_j-\mu_k)^{-1},
        \qquad j=0,\ldots,n,
    \end{equation}
    \begin{equation}\label{5.57}
        \mu_{j,t_r}=\frac{2\widetilde{F}_{r+1}(\mu_j) y(\hat{\mu}_j) }
                    {\mu_j}
         \prod_{ \scriptstyle k=0 \atop \scriptstyle k \neq j}^n
        (\mu_j-\mu_k)^{-1},
        \qquad j=0,\ldots,n,
    \end{equation}
with initial conditions
       \begin{equation}\label{5.58}
         \{\hat{\mu}_j(x_0,t_{0,r})\}_{j=0,\ldots,n}
         \in \mathcal{K}_{n},
       \end{equation}
for some fixed $(x_0,t_{0,r}) \in \Omega_\mu$. The initial value
problem $(\ref{5.57})$, $(\ref{5.58})$ has a unique solution
satisfying
        \begin{equation}\label{5.59}
         \hat{\mu}_j \in C^\infty(\Omega_\mu,\mathcal{K}_{n}),
         \quad j=0,\ldots,n.
        \end{equation}

$\mathrm{(ii)}$
 Suppose that the zeros $\{\nu_l(x,t_r)\}_{l=1,\ldots,n}$
 of $H_n(z,x,t_r)$ remain distinct for $(x,t_r) \in
 \Omega_\nu,$ where $\Omega_\nu \subseteq \mathbb{R}^2$ is open
 and connected. Then,
 $\{\nu_l(x,t_r)\}_{l=1,\ldots,n}$ satisfy the system of
 differential equations,
     \begin{equation}\label{5.60}
        \nu_{l,x}=-2\frac{u_{xx}~y(\hat{\nu}_l)}{h_0 ~ \nu_l}
        \prod_{ \scriptstyle k=1 \atop \scriptstyle k \neq l}^n
        (\nu_l-\nu_k)^{-1},
        \qquad l=1,\ldots,n,
     \end{equation}
     \begin{equation}\label{5.61}
        \nu_{l,t_r}=\frac{2\widetilde{H}_r(\nu_l) y(\hat{\nu}_l) }
                     {h_0~\nu_l}
        \prod_{ \scriptstyle k=1 \atop \scriptstyle k \neq l}^n
        (\nu_l-\nu_k)^{-1},
        \qquad l=1,\ldots,n,
    \end{equation}
with initial conditions
       \begin{equation}\label{5.62}
         \{\hat{\nu}_l(x_0,t_{0,r})\}_{l=1,\ldots,n}
         \in \mathcal{K}_{n},
       \end{equation}
for some fixed $(x_0,t_{0,r}) \in \Omega_\nu$. The initial value
problem $(\ref{5.61})$, $(\ref{5.62})$ has a unique solution
satisfying
        \begin{equation}\label{5.63}
         \hat{\nu}_l \in C^\infty(\Omega_\nu,\mathcal{K}_{n}),
         \quad l=1,\ldots,n.
        \end{equation}
 \end{lem5.4}
\textbf{Proof.}~~ It suffices to focus on (\ref{5.56}), (\ref{5.57})
and (\ref{5.59}), since the proof procedure for (\ref{5.60}),
(\ref{5.61}) and (\ref{5.63}) is similar.

The proof of (\ref{5.56}) has been given in Lemma 3.4. We just
derive (\ref{5.57}). Differentiating on both sides of (\ref{5.6})
with respect to $t_r$ yields
     \begin{equation}\label{5.64}
        F_{n+1,t_r}(\mu_j)=-\mu_{j,t_r}
         \prod_{ \scriptstyle k=0 \atop \scriptstyle k \neq j}^n
        (\mu_j-\mu_k).
     \end{equation}
On the other hand, inserting $z=\mu_j$ into (\ref{5.38}) and
considering (\ref{5.26}), we arrive at
     \begin{equation}\label{5.65}
        F_{n+1,t_r}(\mu_j)=2G_n(\mu_j)\widetilde{F}_{r+1}(\mu_j)
         =2\frac{y(\hat{\mu}_j)}{-\mu_j}\widetilde{F}_{r+1}(\mu_j).
     \end{equation}
Combining (\ref{5.64}) with (\ref{5.65}) leads to (\ref{5.57}). The
proof of smoothness assertion (\ref{5.59}) is analogous to the mCH
case in our latest paper \cite{18}. \quad $\square$ \\

Let us  now present the $t_r$-dependent trace formulas of HS
hierarchy, which are used to construct the algebro-geometric
solutions $u$ in section 6. For simplicity, we just take the
simplest case.

\newtheorem{lem5.5}[lem5.1]{Lemma}
 \begin{lem5.5}
  Assume that $(\ref{2.2})$, $(\ref{5.3})$ and $(\ref{5.4})$ hold.
  Then, we have
    \begin{equation}\label{5.66}
     u=\frac{1}{2}\sum_{j=0}^n \mu_j-\frac{1}{2}\sum_{m=0}^{2n+1}
     E_m.
    \end{equation}
 \end{lem5.5}
\textbf{Proof.}~~The proof is similar to the corresponding
stationary case in Lemma 3.5. \quad $\square$

\section{Time-dependent algebro-geometric solutions}
  In the final section, we extend the results of section 4
  from the stationary HS hierarchy to the time-dependent case.
  In particular,
  we obtain Riemann theta function representations for the
  Baker-Akhiezer function, the meromorphic function $\phi$ and the
  algebro-geometric solutions for the HS hierarchy.

  Let us first consider the asymptotic properties of $\phi$ in the
  time-dependent case.

   \newtheorem{lem6.1}{Lemma}[section]
   \begin{lem6.1}
     Assume that $(\ref{2.2})$,$(\ref{5.3})$ and
     $(\ref{5.4})$ hold. Let $P=(z,y)
    \in \mathcal{K}_n \setminus \{P_{\infty_\pm},P_0\},$ $(x,t_r) \in
    \mathbb{R}^2$. Then, we have
     \begin{equation}\label{6.1}
      \phi(P)\underset{\zeta \rightarrow 0}{=} -u_x+O(\zeta),
      \qquad P \rightarrow P_{\infty_\pm}, \qquad \zeta=z^{-1},
     \end{equation}
     \begin{equation}\label{6.2}
       \phi(P)\underset{\zeta \rightarrow 0}{=} i~a
       \Big(\prod_{m=1}^{2n+1} E_m \Big)^{1/2} f_{n+1}^{-1} \zeta
       +O(\zeta^2),
        \quad P \rightarrow P_0, \quad \zeta=z^{1/2}.
     \end{equation}
   \end{lem6.1}
\textbf{Proof.}~~The proof is identical to the corresponding
stationary case in Lemma 4.1. \quad $\square$ \\

Next, we study the properties of Abel map, which dose not linearize
the divisor
$\mathcal{D}_{\hat{\mu}_0(x,t_r)\underline{\hat{\mu}}(x,t_r)}$ in
the time-dependent HS hierarchy. This is a remarkable difference
between CH, MCH, HS hierarchies and other integrable soliton
equations such as KdV and AKNS hierarchies. For that purpose, we
introduce some notations of symmetric functions.

Let us define
  \begin{equation}\label{6.3}
     \begin{split}
       \mathcal{S}_{k+1}&=\{\underline{l}=(l_1,\ldots,l_{k+1}) \in
       \mathbb{N}_0^{k+1}
        |~l_1 < \cdots < l_{k+1} \leq n\},
        \quad k=0,\ldots, n, \\
       \mathcal{T}_{k+1}^{(j)}&=\{\underline{l}=(l_1,\ldots,l_{k+1}) \in
       \mathcal{S}_{k+1}
      |~ l_m \neq j\}, \quad k=0,\ldots, n-1, ~ j=0, \ldots, n.
     \end{split}
  \end{equation}
The symmetric functions are defined by
    \begin{equation}\label{6.4}
    \Psi_0(\bar{\mu})=1, \quad
    \Psi_{k+1}(\bar{\mu})=(-1)^{k+1} \sum_{\underline{l}\in \mathcal{S}_{k+1}}
        \mu_{l_1}\cdots \mu_{l_{k+1}}, \quad
        k=0,\ldots,n,
    \end{equation}
and
    \begin{equation}\label{6.5}
     \begin{split}
    & \Phi_0^{(j)}(\bar{\mu})=1, \\
    & \Phi_{k+1}^{(j)}(\bar{\mu})=(-1)^{k+1} \sum_{\underline{l}\in \mathcal{T}_{k+1}^{(j)}}
           \mu_{l_1}\cdots \mu_{l_{k+1}}, \\
    & k=0,\ldots,n-1, \quad j=0,\ldots,n,
     \end{split}
     \end{equation}
where $\bar{\mu}=(\mu_0,\ldots,\mu_n) \in \mathbb{C}^{n+1}$. The
properties of $\Psi_{k+1}(\bar{\mu})$ and
$\Phi_{k+1}^{(j)}(\bar{\mu})$ can be found in Appendix E \cite{15}.
Here we freely use these relations.

Moreover, for the HS hierarchy we have
 \footnote{$m \wedge n = \mathrm{min}\{m,n\}$,
 $m \vee n =\mathrm{max} \{m,n\}$}
    \begin{equation}\label{6.6}
      \begin{split}
     & \widehat{F}_{r+1}(\mu_j)= \sum_{s=(r-n)\vee 0}^{r+1}
      \hat{c}_s(\underline{E})
      \Phi_{r+1-s}^{(j)}(\bar{\mu}),
       \\
  & \widetilde{F}_{r+1}(\mu_j)= \sum_{s=0}^{r+1} \tilde{c}_{r+1-s}\widehat{F}_s(\mu_j)
  =
  \sum_{k=0}^{(r+1) \wedge (n+1)} \tilde{d}_{r+1,k}(\underline{E})
  \Phi_k^{(j)}(\bar{\mu}), \quad
    r\in \mathbb{N}_0, ~\tilde{c}_0=1,
     \end{split}
    \end{equation}
where
   \begin{equation}\label{6.7}
     \tilde{d}_{r+1,k}(\underline{E})
     =\sum_{s=0}^{r+1-k} \tilde{c}_{r+1-k-s} \hat{c}_s(\underline{E})
     \qquad k=0,\ldots, r+1 \wedge n+1.
   \end{equation}

\newtheorem{the6.2}[lem6.1]{Theorem}
 \begin{the6.2}
   Assume that $\mathcal{K}_n$ is nonsingular and $(\ref{2.2})$
   holds. Suppose that $\{\hat{\mu}_j\}_{j=0,\ldots,n}$
   satisfies the Dubrovin equations $(\ref{5.56})$, $(\ref{5.57})$
   on $\Omega_\mu$ and remain distinct and $\widetilde{F}_{r+1}(\mu_j)
   \neq 0$ for $(x,t_r)\in \Omega_\mu$, where $\Omega_\mu \subseteq \mathbb{R}^2$
   is open and connected. Introducing the associated divisor
   $\mathcal{D}_{\hat{\mu}_0(x,t_r)\underline{\hat{\mu}}(x,t_r)}$,
   then
   \begin{equation}\label{6.8}
            \partial_x
            \underline{\alpha}_{Q_0}(\mathcal{D}_{\hat{\mu}_0(x,t_r)
            \underline{\hat{\mu}}(x,t_r)})
            =-\frac{2a}{ \Psi_{n+1}(\bar{\mu}(x,t_r))}
             \underline{c}(1),
            \quad    (x,t_r) \in \Omega_\mu,
        \end{equation}
   \begin{eqnarray}\label{6.9}
         \partial_{t_r}
         \underline{\alpha}_{Q_0}(\mathcal{D}_{\hat{\mu}_0(x,t_r)
         \underline{\hat{\mu}}(x,t_r)})
         &=&-\frac{2a}{ \Psi_{n+1}(\bar{\mu}(x,t_r))}
           \\
         & \times &
         \Big( \sum_{k=0}^{(r+1) \wedge (n+1)} \tilde{d}_{r+1,k}(\underline{E})
          \Psi_k(\bar{\mu}(x,t_r)) \Big) \underline{c}(1)
          \nonumber \\
         &+&
         2 a \Big( \sum_{\ell=1 \vee (n+1-r)}^{n+1}
         \tilde{d}_{r+1,n+2-\ell}(\underline{E})
         \underline{c}(\ell)
         \Big),
          \nonumber \\
         &&~~~~~~~~~~~~~~~~~~~~~~~~~~
         (x,t_r) \in \Omega_\mu.
         \nonumber
        \end{eqnarray}
 In particular, the Abel map dose not linearize the divisor
 $\mathcal{D}_{\hat{\mu}_0(x,t_r)\underline{\hat{\mu}}(x,t_r)}$ on $\Omega_\mu$.
 \end{the6.2}
\textbf{Proof.}~~It suffices to prove (\ref{6.9}),  since the proofs
of (\ref{6.8}) has been given in the stationary context of Theorem
4.3. Let us first  give a fundamental identity (E.17) \cite{15},
that is
     \begin{equation}\label{6.10}
     \Phi_{k+1}^{(j)}(\bar{\mu})=\mu_j \Phi_{k}^{(j)}(\bar{\mu})
     +\Psi_{k+1}(\bar{\mu}),
     \quad k=0,\ldots,n, ~
     j=0,\ldots,n.
    \end{equation}
Then, together with (\ref{6.6}) and (\ref{4.16}), we have
    \begin{eqnarray}\label{6.11}
     \frac{\widetilde{F}_{r+1}(\mu_j)}{\mu_j}
     &=&\mu_j^{-1}
     \sum_{m=0}^{(r+1) \wedge (n+1)} \tilde{d}_{r+1,m}(\underline{E})
     \Phi_m^{(j)}(\bar{\mu})
       \\
     &=&
     \mu_j^{-1}
     \sum_{m=0}^{(r+1) \wedge (n+1)} \tilde{d}_{r+1,m}(\underline{E})
     \Big(\mu_j\Phi_{m-1}^{(j)}(\bar{\mu})
     +\Psi_{m}(\bar{\mu})\Big)
       \nonumber \\
     &=&
    \sum_{m=1}^{(r+1) \wedge (n+1)} \tilde{d}_{r+1,m}(\underline{E})
    \Phi_{m-1}^{(j)}(\bar{\mu})
        \nonumber \\
    &&~~~~~
    -\sum_{m=0}^{(r+1) \wedge (n+1)} \tilde{d}_{r+1,m}(\underline{E})
    \Psi_{m}(\bar{\mu})
    \frac{\Phi_{n}^{(j)}(\bar{\mu})}{\Psi_{n+1}(\bar{\mu})}.
     \nonumber
    \end{eqnarray}
So, using (\ref{6.11}), (\ref{5.57}), (E.9), (E.25) and (E.26)
\cite{15}, we obtain
   \begin{eqnarray}
    &&
    \partial_{t_r} \Big(\sum_{j=0}^n \int_{Q_0}^{\hat{\mu}_j}\underline{\omega}\Big)
    =\sum_{j=0}^n \mu_{j,t_r} \sum_{k=1}^n
    \underline{c}(k)\frac{a~\mu_j^{k-1}}{y(\hat{\mu}_j)}
      \nonumber \\
    &&=
    2a\sum_{j=0}^n \sum_{k=1}^n \underline{c}(k)
    \frac{\mu_j^{k-1}}{\prod_{\scriptstyle l=0 \atop \scriptstyle l \neq j}^n (\mu_j-\mu_l)}
    \frac{\widetilde{F}_{r+1}(\mu_j)}{\mu_j}
      \nonumber \\
    &&=
     2a\sum_{j=0}^n \sum_{k=1}^n \underline{c}(k)
    \frac{\mu_j^{k-1}}{\prod_{\scriptstyle l=0 \atop \scriptstyle l \neq j}^n (\mu_j-\mu_l)}
    \Big(
      -\sum_{m=0}^{(r+1) \wedge (n+1)} \tilde{d}_{r+1,m}(\underline{E})
    \Psi_{m}(\bar{\mu})
    \frac{\Phi_{n}^{(j)}(\bar{\mu})}{\Psi_{n+1}(\bar{\mu})}
      \nonumber \\
    &&~~~~~
    +\sum_{m=1}^{(r+1) \wedge (n+1)} \tilde{d}_{r+1,m}(\underline{E})
    \Phi_{m-1}^{(j)}(\bar{\mu}) \Big)
      \nonumber
   \end{eqnarray}
\begin{eqnarray}\label{6.12}
     &=&
     -2a\sum_{m=0}^{(r+1) \wedge (n+1)} \tilde{d}_{r+1,m}(\underline{E})
     \frac{\Psi_{m}(\bar{\mu})}{\Psi_{n+1}(\bar{\mu})}
     \sum_{k=1}^n \sum_{j=0}^n \underline{c}(k)
     (U_{n+1}(\bar{\mu}))_{k,j}(U_{n+1}(\bar{\mu}))_{j,1}^{-1}
       \nonumber \\
     &&
     +2a\sum_{m=1}^{(r+1) \wedge (n+1)} \tilde{d}_{r+1,m}(\underline{E})
     \sum_{k=1}^n \sum_{j=0}^n \underline{c}(k)
     (U_{n+1}(\bar{\mu}))_{k,j}(U_{n+1}(\bar{\mu}))_{j,n-m+2}^{-1}
       \nonumber \\
     &=&
     -\frac{2a}{\Psi_{n+1}(\bar{\mu})}
     \sum_{m=0}^{(r+1) \wedge (n+1)} \tilde{d}_{r+1,m}(\underline{E})
     \Psi_{m}(\bar{\mu}) \underline{c}(1)
       \nonumber \\
     &&
     +2a\sum_{m=1}^{(r+1) \wedge (n+1)} \tilde{d}_{r+1,m}(\underline{E})
     \underline{c}(n-m+2)
        \nonumber \\
     &=&
     -\frac{2a}{\Psi_{n+1}(\bar{\mu})}
     \sum_{m=0}^{(r+1) \wedge (n+1)} \tilde{d}_{r+1,m}(\underline{E})
     \Psi_{m}(\bar{\mu}) \underline{c}(1)
        \nonumber \\
     &&
     +2a\sum_{m=1\vee (n+1-r) }^{ n+1} \tilde{d}_{r+1,n+2-m}(\underline{E})
     \underline{c}(m).
  \end{eqnarray}
Therefore, we complete the proof of (\ref{6.9}). \quad $\square$ \\

The analogous results hold for the corresponding divisor
$\mathcal{D}_{\underline{\hat{\nu}}(x,t_r)}$ associated with
$\phi(P,x,t_r)$.

The following result is a special form of Theorem 6.2, which
provides the constraint condition to linearize the divisor
$\mathcal{D}_{\hat{\mu}_0(x,t_r) \underline{\hat{\mu}}(x,t_r)}$
associated with $\phi(P,x,t_r)$. We recall the definitions of
$\underline{\widehat{B}}_{Q_0}$ and $\underline{\hat{\beta}}_{Q_0}$
in (\ref{4.20}) and (\ref{4.21}).

\newtheorem{the6.3}[lem6.1]{Theorem}
  \begin{the6.3}
   Assume that $(\ref{2.2})$ holds and  the statements of $\{\mu_j\}_{j=0,\ldots,n}$
   in Theorem $6.2$ are true.
   Then,
   \begin{equation}\label{6.13}
        \partial_x \sum_{j=0}^n \int_{Q_0}^{\hat{\mu}_j(x,t_r)} \eta_1
        = -\frac{2a}{\Psi_{n+1}(\bar{\mu}(x,t_r))},
        \qquad (x,t_r) \in \Omega_\mu,
   \end{equation}
   \begin{equation}\label{6.14}
        \partial_x \underline{\hat{\beta}}
        (\mathcal{D}_{\underline{\hat{\mu}}(x,t_r)})
        =
        \begin{cases}
         2a,
         ~~~~~~~~n=1,\\
          2a (0,\ldots,0,1),
          ~~~~~~~~n\geq 2,
        \end{cases}
        \quad  (x,t_r) \in \Omega_\mu,
    \end{equation}
    \begin{eqnarray}\label{6.15}
     \partial_{t_r} \sum_{j=0}^n \int_{Q_0}^{\hat{\mu}_j(x,t_r)} \eta_1
     &=&-\frac{2a}{\Psi_{n+1}(\bar{\mu}(x,t_r))}
     \sum_{k=0}^{(r+1) \wedge (n+1)} \tilde{d}_{r+1,k}(\underline{E})
     \Psi_{k}(\bar{\mu}(x,t_r))
      \nonumber \\
      &+&
     2a \tilde{d}_{r+1,n+1}(\underline{E}) \delta_{n+1,r+1\wedge n+1},
     \quad  (x,t_r) \in \Omega_\mu,
    \end{eqnarray}
    \begin{eqnarray}\label{6.16}
    &&
    \partial_{t_r} \underline{\hat{\beta}}
        (\mathcal{D}_{\underline{\hat{\mu}}(x,t_r)})
        \nonumber \\
        && ~~
        =2a\Big(
     \sum_{s=0}^{r+1} \tilde{c}_{r+1-s}\hat{c}_{s+1-n}(\underline{E}),
     \ldots,
     \sum_{s=0}^{r+1} \tilde{c}_{r+1-s}\hat{c}_{s+1}(\underline{E}),
           \sum_{s=0}^{r+1} \tilde{c}_{r+1-s}\hat{c}_{s}(\underline{E}),
        \Big),
         \nonumber \\
     && ~~~~~~~~~~~~~~~
     \quad \hat{c}_{-l}(\underline{E})=0,~ l\in
     \mathbb{N},
     \quad  (x,t_r) \in \Omega_\mu.
    \end{eqnarray}
 \end{the6.3}
\textbf{Proof.}~~Equations (\ref{6.13}) and (\ref{6.14}) have been
proved in the stationary case in Theorem 4.4. Equations (\ref{6.15})
and (\ref{6.16}) follows from (\ref{6.12}), taking account into
(E.9) \cite{15}. \quad $\square$.\\

Motivated by Theorem 6.2 and Theorem 6.3, the change of variables
    \begin{equation}\label{6.17}
        x \mapsto \tilde{x}= \int^x dx^\prime
        \Big(\frac{2a}{\Psi_{n+1}(\bar{\mu}(x^\prime,t_r))}
        \Big)
    \end{equation}
and
    \begin{eqnarray}\label{6.18}
     t_r \mapsto  \tilde{t}_r &=&\int^{t_r} ds
     \Big(
     \frac{2a}{ \Psi_{n+1}(\bar{\mu}(x,s))}
          \sum_{k=0}^{(r+1) \wedge (n+1)} \tilde{d}_{r+1,k}(\underline{E})
          \Psi_k(\bar{\mu}(x,s))
          \nonumber \\
         && ~~~~~
         -2a \sum_{\ell=1 \vee (n+1-r)}^{n+1}
         \tilde{d}_{r+1,n+2-\ell}(\underline{E})
         \frac{\underline{c}(\ell)}{\underline{c}(1)}
              \Big)
    \end{eqnarray}
linearizes the Abel map
$\underline{A}_{Q_0}(\mathcal{D}_{\hat{\tilde{\mu}}_0(\tilde{x},\tilde{t_r})
\underline{\hat{\tilde{\mu}}}(\tilde{x},\tilde{t}_r)})$,
$\tilde{\mu}_j(\tilde{x},\tilde{t}_r)=\mu_j(x,t_r)$, $j=0,\ldots,n$.
The intricate relation between the variables $(x,t_r)$ and
$(\tilde{x},\tilde{t}_r)$ is detailedly studied in Theorem 6.4.

Next we shall provide an explicit representations of $\phi$ and $u$
in terms of the Riemann theta function associated with
$\mathcal{K}_n$, assuming the affine part of $\mathcal{K}_n$ to be
nonsingular. Since the Abel map fails to linearize the divisor
$\mathcal{D}_{\hat{\mu}_0(x,t_r)\underline{\hat{\mu}}(x,t_r)}$, one
could argue that it suffices to consider the Dubrovin equations
(\ref{5.56})-(\ref{5.57}) and reconstruct $u$ from the trace formula
(\ref{5.66}). By (\ref{4.24})-(\ref{4.32}), one of the principal
results  reads as follows.

\newtheorem{the6.4}[lem6.1]{Theorem}
 \begin{the6.4}
   Suppose that the curve $\mathcal{K}_n$ is nonsingular,
   $(\ref{2.2})$, $(\ref{5.3})$ and
   $(\ref{5.4})$ hold on $\Omega$.
   Let $P=(z,y) \in \mathcal{K}_n \setminus \{P_0\}$,
   and $(x,t_r),(x_0,t_{0,r}) \in \Omega$,  where $\Omega \subseteq
   \mathbb{R}^2$ is open and connected. Moreover, suppose that
   $\mathcal{D}_{\underline{\hat{\mu}}(x,t_r)}$, or
   $\mathcal{D}_{\underline{\hat{\nu}}(x,t_r)}$ is nonspecial for
   $(x,t_r) \in \Omega$. Then, $\phi$ and $u$ have the following
   representations
   \begin{eqnarray}\label{6.19}
    \phi(P,x,t_r)&=&i a \Big(\prod_{m=1}^{2n+1} E_m \Big)^{1/2}
    f_{n+1}^{-1}
    \frac{\theta(\underline{z}(P,\underline{\hat{\nu}}(x,t_r) ))
         \theta(\underline{z}(P_0,\underline{\hat{\mu}}(x,t_r) ))}
     {\theta(\underline{z}(P_0,\underline{\hat{\nu}}(x,t_r) ))
     \theta(\underline{z}(P,\underline{\hat{\mu}}(x,t_r) ))}
      \nonumber \\
     && \times ~
     \mathrm{exp} \Big(d_0-\int_{Q_0}^P \omega_{\hat{\mu}_0(x,t_r) P_0}^{(3)}
     \Big),
   \end{eqnarray}
   \begin{eqnarray}\label{6.20}
    u(x,t_r)&=&-\frac{1}{2}\sum_{m=0}^{2n+1} E_m
             +\frac{1}{2}\sum_{j=1}^n \lambda_j
             \nonumber \\
             &&
    -\frac{1}{2}\sum_{j=1}^n U_j \partial_{\omega_j}
              \mathrm{ln}
              \Big(
    \frac{\theta(\underline{z}(P_{\infty_+},\underline{\hat{\mu}}(x,t_r) )+\underline{\omega})}
         {\theta(\underline{z}(P_{\infty_-},\underline{\hat{\mu}}(x,t_r) )+\underline{\omega})}
              \Big)
              \Big|_{\underline{\omega}=0}
   \end{eqnarray}
Moreover, let $\mu_j$, $j=0,\ldots,n,$ be nonvanishing on $\Omega$.
Then, we have the following constraint
    \begin{eqnarray*}
     &&
     2a(x-x_0)
     +2a(t_r-t_{0,r}) \sum_{s=0}^{r+1} \tilde{c}_{r+1-s} \hat{c}_s(\underline{E})
      \nonumber \\
     &&=
     \Big(-2a \int_{x_0}^x \frac{dx^\prime}{\prod_{k=0}^n \mu_k(x^\prime, t_r)}
          -2a \sum_{k=0}^{(r+1) \wedge (n+1)} \tilde{d}_{r+1,k}(\underline{E})
     \int_{t_{0,r}}^{t_r} \frac{\Psi_k(\bar{\mu}(x_0,s))}{\Psi_{n+1}(\bar{\mu}(x_0,s))}
     ds
     \Big)
     \nonumber \\
     &&~~~~~ \times ~
     \sum_{j=1}^n \Big( \int_{a_j} \tilde{\omega}_{P_{\infty_+}P_{\infty_-}}^{(3)}
     \Big) c_j(1)
  \end{eqnarray*}
  \begin{eqnarray}\label{6.21}
  &&~~~~~+~
     2a(t_r-t_{0,r}) \sum_{\ell=1 \vee (n+1-r) }^{n+1}
     \tilde{d}_{r+1,n+2-\ell}(\underline{E})
     \sum_{j=1}^n \Big( \int_{a_j} \tilde{\omega}_{P_{\infty_+}P_{\infty_-}}^{(3)}
     \Big) c_j(\ell)
      \nonumber \\
     && ~~~~~+~
     \mathrm{ln} \left(
     \frac{\theta(\underline{z}(P_{\infty_+},\underline{\hat{\mu}}(x,t_r) ))
           \theta(\underline{z}(P_{\infty_-},\underline{\hat{\mu}}(x_0,t_{0,r}) ))}
          {\theta(\underline{z}(P_{\infty_-},\underline{\hat{\mu}}(x,t_{r})))
          \theta(\underline{z}(P_{\infty_+},\underline{\hat{\mu}}(x_0,t_{0,r}) ))}
          \right),
    \end{eqnarray}
     $$(x,t_r), (x_0,t_{0,r}) \in \Omega$$
with
  \begin{eqnarray}\label{6.22}
  &&
   \underline{\hat{\alpha}}_{Q_0}(\mathcal{D}_{\hat{\mu}_0(x,t_r)
   \underline{\hat{\mu}}(x,t_r)})
    \nonumber \\
    &&~~
      =\underline{\hat{\alpha}}_{Q_0}(\mathcal{D}_{\hat{\mu}_0(x_0,t_r)
   \underline{\hat{\mu}}(x_0,t_r)})
   -2a\Big(   \int_{x_0}^x
   \frac{dx^\prime}{ \Psi_{n+1}(\bar{\mu}(x^\prime,t_r))} \Big)
             \underline{c}(1)
   \\
   &&~~
   =\underline{\hat{\alpha}}_{Q_0}(\mathcal{D}_{\hat{\mu}_0(x,t_{0,r})
   \underline{\hat{\mu}}(x,t_{0,r})})
     \nonumber \\
   &&~~~~~
   -2a \Big(
       \sum_{k=0}^{(r+1) \wedge (n+1)} \tilde{d}_{r+1,k}(\underline{E})
       \int_{t_{0,r}}^{t_r}
       \frac{\Psi_k(\bar{\mu}(x,s))}{\Psi_{n+1}(\bar{\mu}(x,s))}
       ds \Big)\underline{c}(1)
   \nonumber \\
  &&~~~~~+
       2a(t_r-t_{0,r}) \Big(
       \sum_{\ell=1 \vee (n+1-r) }^{n+1}
       \tilde{d}_{r+1,n+2-\ell}(\underline{E})
       \underline{c}(\ell)
       \Big),
  \end{eqnarray}
       $$(x,t_r), (x_0,t_{0,r}) \in \Omega.$$
 \end{the6.4}
\textbf{Proof.}~~Let us  first assume that $\mu_j$, $j=0,\ldots,n$,
are distinct and nonvanishing on $\widetilde{\Omega}$ and
$\widetilde{F}_{r+1}(\mu_j) \neq 0$ on $\widetilde{\Omega}$,
$j=0,\ldots,n,$ where $\widetilde{\Omega} \subseteq \Omega$. Then,
the representation (\ref{6.19}) for $\phi$ on $\widetilde{\Omega}$
follows by combining (\ref{5.28}), (\ref{6.1}), (\ref{6.2}) and
Theorem A.26 \cite{15}. The representation (\ref{6.20}) for $u$ on
$\widetilde{\Omega}$ follows from the trace formulas (\ref{5.66})
and (F.89) \cite{15}. In fact, since the proofs of (\ref{6.19}) and
(\ref{6.20}) are identical to the corresponding stationary results
in Theorem 4.5, which can be extended line by line to the
time-dependent setting, here we omit the corresponding details. The
constraint (\ref{6.21}) then holds on $\widetilde{\Omega}$ by
combining (\ref{6.13})-(\ref{6.16}), and (F.88) \cite{15}. Equations
(\ref{6.22}) and (6.23) is clear from (\ref{6.8}) and (\ref{6.9}).
The extension of all results from $(x,t_r)\in\widetilde{\Omega}$ to
$(x,t_r)\in\Omega$ then simply follows by the continuity of
$\underline{\alpha}_{Q_0}$ and the hypothesis of
$\mathcal{D}_{\underline{\hat{\mu}}(x,t_r)}$ being nonspecial for
$(x,t_r)\in\Omega$. \quad $\square$

\newtheorem{rem6.5}[lem6.1]{Remark}
 \begin{rem6.5}
  A closer look at Theorem $6.4$ shows that
  $(\ref{6.22})$ and $(6.23)$ equal to
  \begin{eqnarray}\label{6.24}
  \underline{\hat{\alpha}}_{Q_0}(\mathcal{D}_{\hat{\mu}_0(x,t_r)
  \underline{\hat{\mu}}(x,t_r)})
               &=&
  \underline{\hat{\alpha}}_{Q_0}(\mathcal{D}_{\hat{\mu}_0(x_0,t_r)
  \underline{\hat{\mu}}(x_0,t_r)})
       -\underline{c}(1)(\tilde{x}-\tilde{x}_0)
          \\
        &=&
  \underline{\hat{\alpha}}_{Q_0}(\mathcal{D}_{\hat{\mu}_0(x,t_{0,r})
  \underline{\hat{\mu}}(x,t_{0,r})})
        -\underline{c}(1)(\tilde{t}_r-\tilde{t}_{0,r}),
  \end{eqnarray}
  based on the changing of variables $x\mapsto \tilde{x}$ and $t_r \mapsto
  \tilde{t}_r$ in $(\ref{6.17})$ and $(\ref{6.18})$.
  Hence, the Abel map linearizes the divisor
  $\mathcal{D}_{\hat{\mu}_0(x,t_{r})
  \underline{\hat{\mu}}(x,t_{r})}$ on $\Omega$ with
  respect to $\tilde{x}, \tilde{t}_r$.
  This fact reveals that the Abel map does not effect the
  linearization of the divisor
   $\mathcal{D}_{\hat{\mu}_0(x,t_{r})\underline{\hat{\mu}}(x,t_{r})}$
  in the time-dependent HS case.
 \end{rem6.5}

\newtheorem{rem6.6}[lem6.1]{Remark}
  \begin{rem6.6}
    Remark $4.8$ is applicable to the present time-dependent
    context. Moreover, in order to obtain the theta function
    representation of $\psi_j$, $j=1,2,$, one can write $\widetilde{F}_{r+1}$
    in terms of $\Psi_k(\bar{\mu})$ and use $(\ref{5.46})$,
    in analogy to the stationary case studied in Remark $4.9$. Here
    we skip the corresponding details.
  \end{rem6.6}

Let us end this section by providing another principle result about
algebro-geometric initial value problem of HS hierarchy. We will
show that the solvability of the Dubrovin equations (\ref{5.56}) and
(\ref{5.57}) on $\Omega_\mu \subseteq \mathbb{R}^2 $ in fact implies
(\ref{5.3}) and (\ref{5.4}) on $\Omega_\mu$. As pointed out in
Remark 4.12, this amounts to solving the time-dependent
algebro-geometric initial value problem (\ref{5.1}) and (5.2) on
$\Omega_\mu$. Recalling definition of $\widetilde{F}_{r+1} (\mu_j)$
introduced in (\ref{6.6}), then we may present the following result.

\newtheorem{the6.7}[lem6.1]{Theorem}
   \begin{the6.7}
     Assume that $(\ref{2.2})$ holds and
     $\{\hat{\mu}_j\}_{j=0,\ldots,n}$ satisfies the Dubrovin
     equations $(\ref{5.56})$ and $(\ref{5.57})$ on $\Omega_\mu$ and
     remain distinct and nonzero for $(x,t_r) \in \Omega_\mu$, where
     $\Omega_\mu \subseteq \mathbb{R}^2 $ is open and connected.
     Moreover, suppose that $\widetilde{F}_{r+1} (\mu_j)$ in
     $(\ref{5.57})$ expressed in terms of $\mu_k$, $k=0,\ldots,n$ by
     $(\ref{6.6})$. Then $u \in C^\infty(\Omega_\mu)$ defined by
       \begin{equation}\label{6.26}
   u=-\frac{1}{2}\sum_{m=0}^{2n+1} E_m +\frac{1}{2}\sum_{j=0}^n \mu_j,
       \end{equation}
    satisfies the $r$th HS equation $(\ref{5.1})$, that is,
       \begin{equation}\label{6.27}
        \mathrm{HS}_r(u)=0
        \quad \textrm{on $\Omega_\mu$},
       \end{equation}
     with initial values satisfying the $n$th stationary HS equation $(5.2)$.
   \end{the6.7}
\textbf{Proof.}~~Given the
  solutions $\hat{\mu}_j=(\mu_j,y(\hat{\mu}_j))\in
  C^\infty(\Omega_\mu,\mathcal {K}_n),$ $j=0,\cdots,n$
  of (\ref{5.56}) and (\ref{5.57}), we
  introduce polynomials $F_{n+1}, G_n,$ and $H_n$
  on $\Omega_\mu$, which are exactly the same as in Theorem 4.10 in the
  stationary case
      \begin{eqnarray}
       &&
        F_{n+1}(z)= \prod_{j=0}^n (z-\mu_j), \\
       &&
        G_n(z)=\frac{1}{2}F_{n+1,x}(z), \\
       &&
       zG_{n,x}(z)=-H_n(z)-u_{xx}F_{n+1}(z),
          \\
       &&
       H_{n,x}(z)=2u_{xx}G_n(z), \\
       &&
       R_{2n+2}(z)=z^2G_n^2(z)+zF_{n+1}(z)H_n(z),
     \end{eqnarray}
where $t_r$ is treated as a parameter. Hence let us focus on
the proof of (\ref{5.1}).\\
Let us denote the polynomial $\widetilde{G}_r$ and $\widetilde{H}_r$
of degree $r$ by
     \begin{equation}\label{6.33}
        \widetilde{G}_r(z)=\frac{1}{2}\widetilde{F}_{r+1,x}(z)
        \quad \textrm{on $ \mathbb{C} \times \Omega_\mu$},
     \end{equation}
     \begin{equation}\label{6.34}
     \widetilde{H}_r(z)=-z\widetilde{G}_{r,x}(z)-u_{xx}\widetilde{F}_{r+1}(z)
     \quad \textrm{on $ \mathbb{C} \times \Omega_\mu$},
     \end{equation}
respectively. Next we want to establish
    \begin{equation}\label{6.35}
        F_{n+1,t_r}(z)=2(G_n(z)\widetilde{F}_{r+1}(z)-F_{n+1}(z)\widetilde{G}_r(z))
        \quad \textrm{on $ \mathbb{C} \times \Omega_\mu$}.
    \end{equation}
One computes from (\ref{5.56}) and (\ref{5.57}) that
   \begin{equation}\label{6.36}
   F_{n+1,x}(z)=-F_{n+1}(z) \sum_{j=0}^n \mu_{j,x} (z-\mu_j)^{-1},
   \end{equation}
   \begin{equation}\label{6.37}
   F_{n+1,t_r}(z)=-F_{n+1}(z) \sum_{j=0}^n
   \widetilde{F}_{r+1}(\mu_j) \mu_{j,x} (z-\mu_j)^{-1}.
   \end{equation}
Using (6.29) and (\ref{6.33}) one concludes that (\ref{6.35}) is
equivalent to
    \begin{equation}\label{6.38}
    \widetilde{F}_{r+1,x}(z)=\sum_{j=0}^n (\widetilde{F}_{r+1}(\mu_j)
    -\widetilde{F}_{r+1}(z)) \mu_{j,x} (z-\mu_j)^{-1}.
    \end{equation}
Equation (\ref{6.38}) has been proved in Lemma F.9 \cite{15}. Hence
this in turn proves (\ref{6.35}). \\
Next, differentiating (6.29) with respect to $t_r$  yields
    \begin{equation}\label{6.39}
    F_{n+1,xt_r}=2G_{n,t_r}.
    \end{equation}
On the other hand, the derivative of (\ref{6.35}) with respect to
$x$, taking account into (6.29), (6.30) and (\ref{6.33}), we obtain
    \begin{eqnarray}\label{6.40}
     F_{n+1,t_rx}&=&-2z^{-1}H_n\widetilde{F}_{r+1}
     -2z^{-1}u_{xx}F_{n+1}\widetilde{F}_{r+1}+2G_n\widetilde{F}_{r+1,x}
       \nonumber \\
     &&
     -2\widetilde{G}_{r,x}F_{n+1}-4\widetilde{G}_rG_n.
    \end{eqnarray}
Combining (\ref{6.34}), (\ref{6.39}) and (\ref{6.40}) we conclude
    \begin{equation}\label{6.41}
     zG_{n,t_r}(z)=\widetilde{H}_r(z)F_{n+1}(z)-H_n(z)\widetilde{F}_{r+1}(z)
      \quad \textrm{on $ \mathbb{C} \times \Omega_\mu$. }
    \end{equation}
Next, differentiating (6.32) with respect to $t_r$, inserting the
expressions (\ref{6.35}) and (\ref{6.41}) for $F_{n+1,t_r}$ and
$G_{n,t_r}$, respectively, we obtain
    \begin{equation}\label{6.42}
        H_{n,t_r}(z)=2(H_n(z)\widetilde{G}_r(z)-G_n(z)\widetilde{H}_r(z))
        \quad \textrm{on $ \mathbb{C} \times \Omega_\mu$. }
    \end{equation}
Finally, taking the derivative of (\ref{6.41}) with respect to $x$
and inserting (6.29), (6.31) and (\ref{6.33}) for $F_{n+1,x}$,
$H_{n,x}$ and $\widetilde{F}_{r+1,x}$, respectively, yields
     \begin{equation}\label{6.43}
     zG_{n,t_rx}= F_{n+1}\widetilde{H}_{r,x}+2G_n\widetilde{H}_r
      -2u_{xx}G_n\widetilde{F}_{r+1}
         -2H_n\widetilde{G}_r.
    \end{equation}
On the other hand, differentiating (6.30) with respect to $t_r$,
using (\ref{6.35}) and (\ref{6.42}) for $F_{n+1,t_r}$ and
$H_{n,t_r}$, respectively, leads to
   \begin{equation}\label{6.44}
    zG_{n,xt_r}=2G_n\widetilde{H}_r-2H_n\widetilde{G}_r
    -u_{xxt_r}F_{n+1}-2u_{xx}(G_n\widetilde{F}_{r+1}-\widetilde{G}_{r}F_{n+1})
   \end{equation}
Hence, combining (\ref{6.43}) and (\ref{6.44}) then yields
   \begin{equation}\label{6.45}
   -u_{xxt_r}-\widetilde{H}_{r,x}+2u_{xx}\widetilde{G}_r=0.
   \end{equation}
Thus we proved (5.12)-(5.17) and (5.38)-(5.40) on $ \mathbb{C}
\times \Omega_\mu$ and hence conclude that $u$ satisfies the $r$th
HS equation (\ref{5.1}) with initial values satisfying the $n$th
stationary HS equation (5.2)  on $ \mathbb{C} \times \Omega_\mu$.
\quad $\square$

\newtheorem{rem6.8}[lem6.1]{Remark}
     \begin{rem6.8}
       The result in Theorem $6.7$ is presented in terms of
       $u$ and $\{\mu_j\}_{j=0,\cdots,n}$, but of course
        one can provide the analogous result
       in terms of $u$ and $\{\nu_l\}_{l=1,\cdots,n}$.
     \end{rem6.8}
The analog of Remark 4.13 directly extends to the current
time-dependent HS hierarchy.

\section*{Acknowledgments}
YH and PZ are very grateful to Professor F.Gesztesy for his helps
about CH solutions and relativistic Toda solutions. YH would also
like to thank Professor E.G. Reyes for many valuable suggestions.
This work was supported by grants from the National Science
Foundation of China (Project No.10971031), and the Shanghai Shuguang
Tracking Project (Project No.08G\\G01).


\begin{thebibliography}{99}
\bibitem{1}J.K. Hunter, R. Saxton, Dynamics of director fields,
 SIAM J. Appl. Math. 51 (1991), 1498--1521.
\bibitem{2}J.K. Hunter, Y.X. Zheng, On a completely integrable nonlinear
hyperbolic variational equation, Physica D. 79 (1994) 361--386.
\bibitem{3}J.K. Hunter, Y.X. Zheng, On a nonlinear hyperbolic variational
equation: I, global existence of weak solutions, Arch. Ration. Mech.
Anal. 129 (1995) 305每353.
\bibitem{4} M.J. Ablowitz, D.J. Kaup, A.C. Newell and H. Segur, The inverse
 scattering transform--Fourier analysis for nonlinear
 problems, Stud.Appl.Math. 53 (1974) 249--315.
\bibitem{5} R. Camassa, D.D. Holm, An integrable shallow water
equation with peaked solitons, Phys.Rev.Lett. 71 (1993) 1661--1664.
\bibitem{6} R. Camassa, D.D. Holm, J.M. Hyman, A new integrable shallow
water equation, Adv.Appl.Mech. 31 (1994) 1--33.
\bibitem{7} E.D. Belokolos, A.I. Bobenko, V.Z. Enol'skii, A.R. Its,
and V.B. Matveev, Algebro-Geometric Approach to Nolinear Integrable
Equations, Springer, Berlin, 1994.
\bibitem{8} S.P. Novikov, S.V. Manakov, L.P. Pitaevskii, V.E.
Zakharov, Theory of Solitons, the Inverse Scattering Methods,
Concultants Bureau, New York, 1984.
\bibitem{9} B.A. Dubrovin, Completely integrable Hamiltonian systems
associated with matrix operators and Abelian varieties,
Funct.Anal.Appl. 11 (1977) 265--277.
\bibitem{10} B.A. Dubrovin, Theta functions and nonlinear equations,
Russian Math.Surveys. 36 (1981) 11--80.
\bibitem{11} B.A. Dubrovin, Matrix finite-gap operators,
Revs.Sci.Tech. 23 (1983) 33--78.
\bibitem{12} F. Gesztesy and R. Ratneseelan, An alternative approach
to algebro-geometric solutions of the AKNS hierarchy, Rev.Math.Phys.
10 (1998) 345--391.
\bibitem{13}  F. Gesztesy and H. Holden, Algebro-geometric solutions
of the Camassa-Holm hierarchy, Rev.Mat.Iberoam. 19 (2003) 73--142.
\bibitem{14} F. Gesztesy and H. Holden, Real-valued algebro-geometric
solutions of the Camassa-Holm hierarchy, Phil.Trans.R.Soc.A. 366
(2008) 1025--1054.
\bibitem{15} F. Gesztesy and H. Holden, Soliton Equations and their
Algebro-Geometric Solutions, Cambridge University Press, Cambridge,
2003.
\bibitem{16} Y. Hou and E.G. Fan, Algebro-geometric solutions of
Gerdjikov-Ivanov hierarchy, preprint, (2011).
\bibitem{17} Y. Hou, P. Zhao, E.G. Fan and Z.J. Qiao, The global solutions
 of algebro-geometric type for Degasperis-Procesi hierarchy,
 preprint, arXiv: 1204.2140 (2012).
\bibitem{18} Y. Hou, E.G. Fan and Z.J. Qiao, The algebro-geometric solutions
for the modified Camassa-Holm hierarchy, preprint, arXiv: 1205.6062
(2012).
\bibitem{19}R. Beals, D.H. Sattinger, J. Szmigielski, Inverse scattering
solutions of the Hunter每Saxton equation, Appl. Anal. 78 (2001)
255--269.
\bibitem{20}E.G. Reyes, The soliton content of the Camassa每Holm and
Hunter每Saxton equations,in: A.G. Nikitin, V.M. Boyko, R.O. Popovych
(Eds.), Proceedings of the Fourth International Conference on
Symmetry in Nonlinear Mathematical Physics, in: Proceedings of the
Institute of Mathematics of the NAS of Ukraine, vol. 43, Kyiv, 2002,
pp. 201--208.
\bibitem{21}E.G. Reyes, Pseudo-potentials, nonlocal symmetries,
and integrability of some shallow water equations, Selecta Math.
(N.S.) 12 (2006) 241--270.
\bibitem{22}B. Khesin, G. Misio{\l}ek, Euler equations on homogeneous spaces
and Virasoro orbits, Adv. Math. 176 (2003) 116-144.
\bibitem{23}J. Lenells, Weak geodesic flow and global solutions of
the Hunter每Saxton equation, Discrete Contin. Dyn. Syst. 18 (2007)
643--656.
\bibitem{24}J. Lenells, The Hunter-Saxton equation describes the
geodesic flow on a sphere, J. Geom. Phys. 57 (2007) 2049--2064.
\bibitem{25}A. Bressan, A. Constantin, Global solutions of the
 Hunter-Saxton equation, SIAM J. Math. Anal. 37 (2005) 996--1026.
\bibitem{26} A. Bressan, H. Holden and X. Raynaud, Lipschitz metric
for the Hunter-Saxton equation, J. Math. Pure. Appl. 94 (2010)
68--92.
\bibitem{27}Z. Yin, On the structure of solutions to the
periodic Hunter-Saxton equation, SIAM J. Math. Anal. 36 (2004)
272--283.
\bibitem{28} G.L. Gui, Y. Liu and M. Zhu, On the wave-breaking phenomena
and global existence for the generalized periodic Camassa-Holm
equation, Int. Math. Res. Notices. 10 (2011) 1--46.
\bibitem{29} O.I. Morozov, Contact equivalence of the generalized
Hunter-Saxton equation and the Euler-Poisson equation. Preprint
math-ph/0406016.
\bibitem{30} S. Sakovich, On a Whitham-type equation, Symmetry,
Integrability. Geom: Methods. Appl. (SIGMA) 5 (2009) 1--7.
\bibitem{31} A.S. Fokas, B. Fuchssteiner, Symplectic structures,
their B$\mathrm{\ddot{a}}$cklund transformation and hereditary
symmetries. Phys. D. 4 (1981) 47--66.
\bibitem{32}P. Rosenau, Nonlinear dispersion and compact structures,
 Phys. Rev. Lett. 73 (1994) 737--1741.
\bibitem{33} V. B. Martveev and M.I. Yavor, Solutions presque
p$\mathrm{\acute{e}}$riodiques et $\mathrm{\grave{a}}$ $N$-solitons
de l'$\mathrm{\acute{e}}$quation hydrodynamique non
lin$\mathrm{\acute{e}}$aire de Kaup, Ann.Inst.
H.Poincar$\mathrm{\acute{e}}$ Sect. A 31 (1979) 25--41.

\end{thebibliography}
\end{document}